\documentclass[10pt,journal,compsoc]{IEEEtran}

\usepackage[colorlinks]{hyperref}
\usepackage{kotex}
\usepackage{enumitem}
\usepackage[T1]{fontenc}

\ifCLASSOPTIONcompsoc
  \usepackage[nocompress]{cite}
\else
  \usepackage{cite}
\fi

\ifCLASSINFOpdf
   \usepackage[pdftex]{graphicx}
   \graphicspath{{../pdf/}{../jpeg/}{./image}}
   \DeclareGraphicsExtensions{.pdf,.jpeg,.png,.jpg}
\else
   \usepackage[dvips]{graphicx}
   \graphicspath{{../eps/}}
\fi

\usepackage[dvipsnames]{xcolor}
\definecolor{amethyst}{rgb}{0.6, 0.4, 0.8}

\newcommand{\resized}[1]{\vcenter{\hbox{\includegraphics[height=1em]{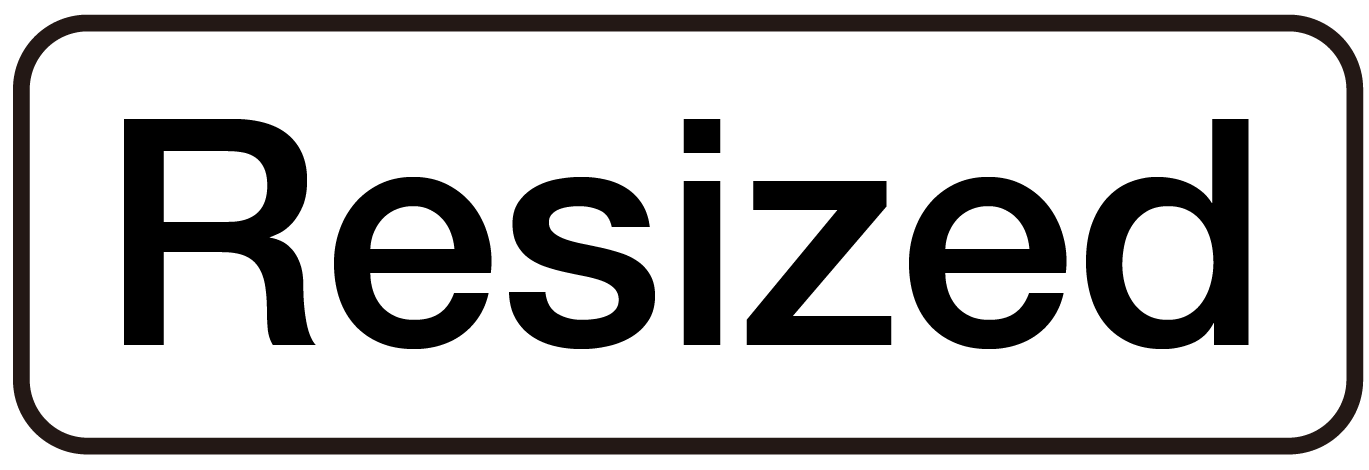}}}}
\newcommand{\gnrd}[1]{\vcenter{\hbox{\includegraphics[height=1em]{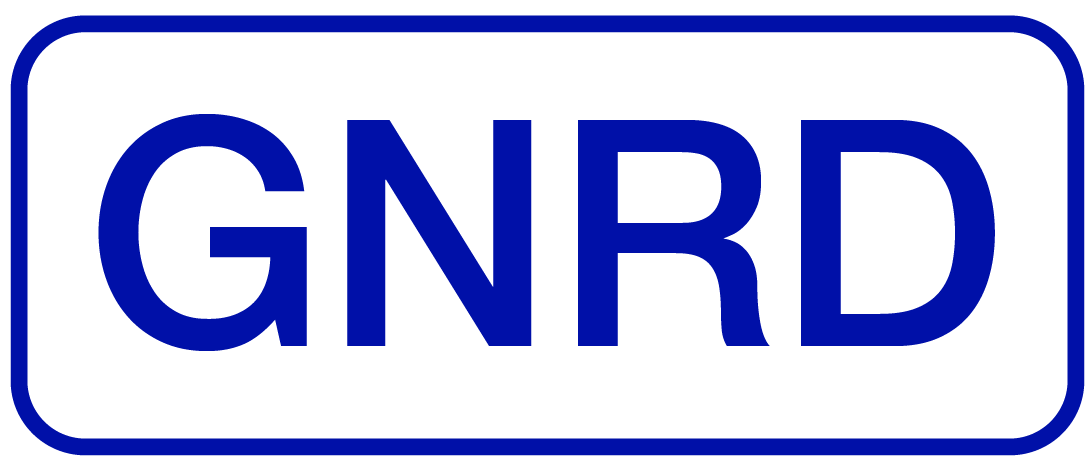}}}}
\newcommand{\highlight}[1]{\vcenter{\hbox{\includegraphics[height=1em]{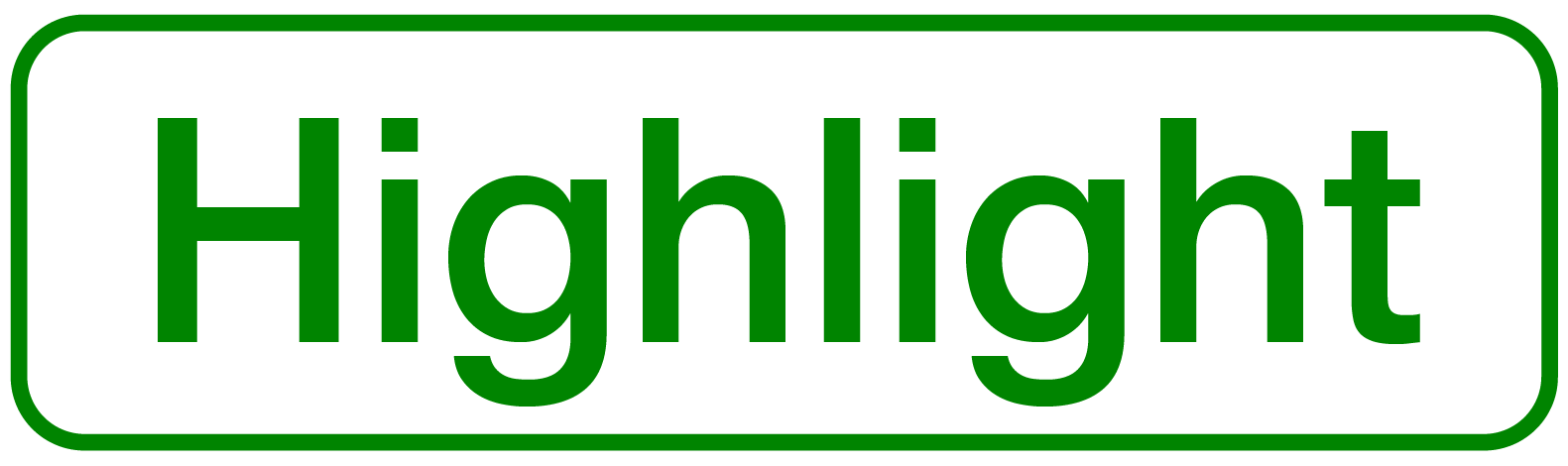}}}}
\newcommand{\hro}[1]{\vcenter{\hbox{\includegraphics[height=1em]{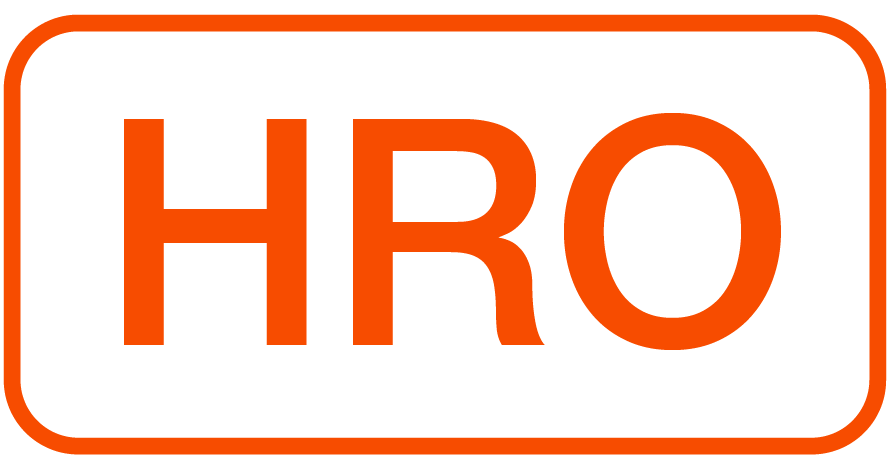}}}}

\newcommand{\resizedt}[1]{\vcenter{\hbox{\includegraphics[height=1em]{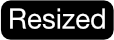}}}}
\newcommand{\gnrdt}[1]{\vcenter{\hbox{\includegraphics[height=1em]{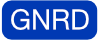}}}}
\newcommand{\highlightt}[1]{\vcenter{\hbox{\includegraphics[height=1em]{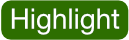}}}}
\newcommand{\hrot}[1]{\vcenter{\hbox{\includegraphics[height=1em]{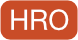}}}}

\begin{document}

\title{Towards Visualization Thumbnail Designs that Entice Reading Data-driven Articles}

\author{Hwiyeon~Kim,
        Joohee~Kim,
        Yunha~Han,
        Hwajung~Hong,
        Oh-Sang~Kwon,
        Young-Woo~Park, 
        Niklas~Elmqvist, 
        Sungahn~Ko, 
        and Bum~Chul~Kwon
\IEEEcompsocitemizethanks{
\IEEEcompsocthanksitem H.\ Kim, J.\ Kim, Y.\ Han, O.\ Kwon, Y.\ Park, and S.\ Ko (corresponding author) are with UNIST, Ulsan, South Korea.
\IEEEcompsocthanksitem H.\ Hong is with KAIST, Daejeon, South Korea.
\IEEEcompsocthanksitem N.\ Elmqvist is with the University of Maryland, College Park, MD, USA
\IEEEcompsocthanksitem B.C.\ Kwon is with IBM Research, Cambridge, MA, USA}
}

\markboth{To appear in IEEE Transactions on Visualization and Computer Graphics}%
{Kim \MakeLowercase{\textit{et al.}}: Thumbnails for Data Stories: A Survey of Current Practices}

\IEEEtitleabstractindextext{%
\begin{abstract}
As online news increasingly include data journalism, there is a corresponding increase in the incorporation of visualization in article thumbnail images.
    However, little research exists on the design rationale for visualization thumbnails, such as resizing, cropping, simplifying, and embellishing charts that appear within the body of the associated article. 
    Therefore, in this paper we aim to understand these design choices and determine what makes a visualization thumbnail inviting and interpretable. 
    To this end, we first survey visualization thumbnails collected online and discuss visualization thumbnail practices with data journalists and news graphics designers. 
    Based on the survey and discussion results, we then define a design space for visualization thumbnails and conduct a user study with four types of visualization thumbnails derived from the design space. 
    The study results indicate that different chart components play different roles in attracting reader attention and enhancing reader understandability of the visualization thumbnails.
    We also find various thumbnail design strategies for effectively combining the charts' components, such as a data summary with highlights and data labels, and a visual legend with text labels and Human Recognizable Objects (HROs), into thumbnails.
    Ultimately, we distill our findings into design implications that allow effective visualization thumbnail designs for data-rich news articles. 
    Our work can thus be seen as a first step toward providing structured guidance on how to design compelling thumbnails for data stories.
\end{abstract}

}

\maketitle
 
\IEEEdisplaynontitleabstractindextext
\IEEEpeerreviewmaketitle

\ifCLASSOPTIONcompsoc
\IEEEraisesectionheading{
\section{Introduction}
\label{sec:introduction}
}

\IEEEPARstart{T}{humbnails} are small and static images that accompany titles and bylines in documents~\cite{Li08, Aula10, Song16}. 
News organizations have extensively used thumbnails for their online news articles to induce reader attention and attract clicks.
As \textit{data stories}---where visualization plays an important role---grow in popularity, news organizations are starting to publish thumbnails that incorporate visualization into the thumbnails.
These \textit{visualization thumbnails} are different from conventional thumbnails because they have great potential to both convey an article's main message using its embedded visualization as well as entice the reader to click on the article link based on the appeal of the displayed data. 

Several news media outlets have begun to adopt such visualization thumbnails. 
For example, \textit{The New York Times} presents many visualization thumbnails in its data story dedicated section ``The Upshot."
\textit{FiveThirtyEight} showcases many visualization thumbnails that can effectively catch the reader's attention. 
Although many types of visualization thumbnails have been designed, few studies have investigated the variety of design choices.
As a result, many questions remain unanswered about optimal visualization thumbnail design.
For example, ``The Upshot'' and \textit{The Economist} tend to use thumbnails that are resized from visualizations embedded in the article.
Meanwhile, \textit{The Pudding}, \textit{FiveThirtyEight}, and \textit{The Wall Street Journal} often design visualization thumbnails with editing choices different from the source visualizations in the article by adding annotations or removing axes.
However, it is unclear which factors influence both \textit{interpretability}---helping readers understand the article from the thumbnail alone---and \textit{appeal}---inducing the reader to click to read the article. 

In this research, we aim to investigate visualization thumbnail designs that support both of these affordances.
To this end, we first survey existing visualization thumbnails from news organizations and find different types of thumbnail designs (Section~\ref{sec_surveythumbnails}).
Then we interview six industry practitioners to understand their thumbnail design strategies and intentions (Section~\ref{sec_interview}).  
With the survey and interview results, we conduct a user study ($n=161$) using 16 thumbnails that we carefully design with industry practitioners to investigate (RQ1) visualization thumbnail designs that readers want to see, and (RQ2) the roles of thumbnail components (Section~\ref{sec_userStudy}). 
The results indicate that readers want visualization thumbnails that quickly capture their attention and effectively convey the articles' main point (Section~\ref{sec_result}). 
We report the lessons learned and design implications derived from the study and discuss open research areas on visualization thumbnails (Section~\ref{sec_DesignImplications}). 
To our knowledge, this is the first attempt to explore the design space, component roles, and readers' choice of visualization thumbnails with evaluation.

This paper is a significantly extended version of a IEEE VIS 2019 short paper surveying visualization thumbnails for data stories~\cite{Kim19}.
Compared to this earlier report, this paper includes empirical evaluation involving 161 crowdworkers as well as our corresponding analysis and findings.
The supplemental material for this journal version of the work can be found on OSF: \url{https://osf.io/khgw2/}

The main contributions of this work are as follows:
(1) Extracting the key components of visualization thumbnail designs by surveying in-the-wild visualization thumbnails;
(2) identifying reader preferences for visualization thumbnail designs and their rationale for their preferences via a crowdsourced user study;
and (3) reporting lessons, design implications, and open research areas for visualization thumbnails.
\section{Related work}

We situate our work in relation to thumbnail design and to the use of visualization for communication, particularly in data journalism. 

\subsection{Thumbnail Design}

Prior research has shown that thumbnails help people \textit{locate} and \textit{rediscover} content during web searches, media browsing, and sensemaking.
For example, when searching the web, thumbnails paired with titles, text snippets, and URLs help people find articles of interest online~\cite{Aula10, Dziadosz02, Woodruff01}. 
Aula et al.~\cite{Aula10} show that image thumbnails add information about the relevance of the web page compared to the text summary.
Thumbnails also appear when browsing other types of media such as file systems~\cite{Robertson98}, documents~\cite{Cockburn06}, and videos~\cite{Matejka13}.
When an email contains a link to a video file, Topkara et al.~\cite{Topkara12} finds that more people click on the link displayed with the thumbnail. 
Previous work on visualization and visual analysis has incorporated thumbnails into the sensemaking process as a means of leveraging the unique advantages of spatial memory~\cite{Nguyen16, Yoghourdjian18, Heer08}.

In this study, we investigate the role thumbnails play as people browse online news articles in the context of data journalism.
On online news webpages, thumbnails of articles must compete for the readers' attention with one another and with other content.
There are several factors that affect the interest and ultimate usefulness of a thumbnail, such as the size of the thumbnail and the text it contains.
For example, Kaasten et al.~\cite{Kaasten02}  conclude that thumbnails must be larger than $96\times96$ pixels to trigger recognition among returning readers to pages with multiple thumbnails.
Woodruff et al.~\cite{Woodruff01} also find that enhanced versions of thumbnails including keywords have better search performance than text summaries or image thumbnails.

Previous studies have been conducted to automatically generate image thumbnails by cropping, reducing, or selecting salient parts of photography included in articles~\cite{Li08, Suh03, Bylinskii17}.
Song et al.~\cite{Song16} present an algorithm for selecting the most salient and evocative thumbnail among image candidates to represent a video.
Compared to this prior art, our study is the first to study visualization images, as well as the first step towards the automatic creation of visualization thumbnails based on a design space for data journalism.

\subsection{Storytelling in Thumbnails}

Thumbnails can be a key player in visual communication and storytelling~\cite{Agrawala11, Hullman11}, as their innate goal is to effectively convey a story in a concise visual format. 
But as they are usually given a smaller space budget, there have been many discussions on ``how much is much'' (i.e., a debate between minimalism and chartjunk).
Many visualization researchers have conventionally supported minimalism in visualization, stating ``less is more,'' as they believe that objective characteristics in visualization improve visualization performance. 
Edward Tufte~\cite{Tufte01}, one of the most vocal proponents of maximizing the ``data-to-ink'' ratio, argues that ``ink that fails to depict statistical information does not have much interest to the viewer.'' 

There are researchers who argue against minimalism in visualization, stating that complexity (e.g., chartjunk~\cite{Bateman10}) is not a bad thing and could make a design more attractive~\cite{Norman07}.
Many experiments have been performed over the years, but results seem to support that ``chart junk'' or embellished graphics (e.g., Nigel Holmes' work) is not a bad thing.
Kelly~\cite{Kelly88} conducted such an experiment and found that 120 participants made similar errors in comprehension and search tasks with the infographics data (collected from USA Today), regardless of data-ink ratios.
Other researchers also observe in their experiments that visual embellishment does not have a significant impact on visualization performance~\cite{Skau15, Moere12}.
Rather, visual embellishment even results in better results in supporting human recall ability~\cite{Bateman10, Borgo12, Borkin13, Borkin16} and qualitative dimensions (e.g., aesthetics and preference)~\cite{Inbar07, Hill17}. 
Due to the importance of annotations (i.e., explanation text in this work) in story-telling, which is also shown in our study, 
Ren et al.~\cite{Ren17} propose a visual tool for efficiently producing annotations, which can be greatly useful for making thumbnails with ample context.

Annotations or other visual components may carry a bias or slant~\cite{Kong18}. 
Whether intended or not~\cite{Pandey15}, a spotted bias on thumbnail's annotations negatively impacts an article's credibility and legitimacy.
Kong and Agrawala~\cite{Kong12} show that placing additional layered information could aid chart reading.
We report how visual components on visualization thumbnails can impact article reading and credibility.

\subsection{Visualization in Data Journalism}

With data journalism on the rise, visualization is becoming pervasive in news media~\cite{stolper2018}.
Different intentions for visualization communication often lead to different design choices~\cite{kosara2016presentation}, encountering substantial use of graphical and text-based annotation~\cite{Ren17}.
We see that some visualizations incorporate human-recognizable objects~\cite{haroz2015isotype}, such as icons and logos, to help people understand the data.
Other graphics that are not directly relevant to data~\cite{Byrne16}, such as portraits and illustrations, are also incorporated into visualizations and considered embellishments~\cite{Bateman10, Skau15} to enhance the meaning of the chart.
These graphical annotations are considered aesthetic design choices~\cite{Moere12} that make positive first impressions~\cite{Harrison15} on the reader.
However, there is still debate as to whether these decorations have a positive effect on the informativeness of charts.
While some studies have shown that embellishments can increase comprehension~\cite{Kong12} and memory~\cite{Borgo12, Borkin13, Borkin16}, Hullman and Adar~\cite{Hullman11} argue that some embellishments make charts more difficult to interpret.
Andry et al.~\cite{Andry21} quantitatively measure people's understanding of chart embellishments and show that there is a positive effect within limited boundaries.

Since online news articles must hook users in a limited time~\cite{Amini2018DDS}, the design of the first impression~\cite{Harrison15} is vital.
Transparency of visualization is consistently emphasized as a challenge in data journalism~\cite{Kennedy20}.
Some practices of generating slanted titles~\cite{Kong18} or designing deceptive visualizations~\cite{Pandey15} can bias and often mislead readers.
Designing thumbnails with visualizations is also important because the thumbnails represent their articles and provide first impressions for the articles.
Therefore, there is a need for research on how to design attractive and effective thumbnails that can be easily interpreted without misleading the reader.
Building on the above understanding, we further explore the current practices of visualization thumbnail design.

\section{Visualization Thumbnail Design Practice}
\label{sec_survey}

To our knowledge, prior research has yet to examine the design of thumbnails in data journalism, and particularly those that we refer to as \textit{visualization thumbnails}.
To better understand current practices in visualization thumbnail design, we collected examples from data journalism outlets and conducted semi-structured interviews with six graphic designers working for news websites.

\subsection{A Survey of Visualization Thumbnails}
\label{sec_surveythumbnails}

We began by collecting news articles published between November 1, 2018 and December 31, 2018 from online news organizations reputed for their data journalism, the \textit{Pew Research Center} (Pew.), \textit{The Economist} (Eco.),~\textit{The New York Times} (NYT), including The Upshot and DealBook, \textit{FiveThirtyEight} (538), \textit{The Wall Street Journal} (WSJ), \textit{First Tuesday Journal} (1st), and \textit{Bloomberg News} (BBG).  
We concentrated on articles relating to politics and economics, as these topics tend to use visualization to a greater degree than others covered by these news outlets. 
This initial corpus contained \textbf{139} articles that include visualizations within the article body.
Among these, 48 articles did not feature visualizations in the article's thumbnail, instead opting for photographs and other visual imagery aside from visualization.
Of the remaining \textbf{91}, we decided to focus on basic charts, such as bar and line charts and scatterplots, because they are most common in news media.
Thus we excluded 24 articles whose charts are not axis-based (e.g., maps) or are infrequently used (e.g., pictogram charts, Sankey diagrams).

Among the remaining \textbf{67} articles, 39 included a visualization used in the body of the article as its thumbnail, reducing its size or cropping it. 
Thumbnails for the remaining 28 articles modified a visualization used in the body of the article in some way (e.g., omitting axes).
Examining these 67 thumbnails further, we codified 1) which visualization components were modified (e.g., an axis); and 2) how components were modified (e.g., omitted).
Three authors of this paper independently codified these components and later merged their codes in an iterative discussion, arriving at 96\% agreement (Fleiss' Kappa $=0.75$).

To label the components, we initially considered using existing classifications, namely Borkin et al.'s classification of visualization components~\cite{Borkin16}, Byrne et al.'s~\cite{Byrne16} distinction between graphical and figurative components~\cite{Byrne16}, and Ren et al.'s~\cite{Ren17} classification of annotation.
We realized that these classifications were insufficient in isolation in terms of capturing all aspects of visualization thumbnail design. 
For example, Borkin et al.'s classification defines ``text'' as ``any text in the image,'' so axis titles, annotations, and captions fall under the same category.
Meanwhile, Byrne et al.'s classification provided broad categories for coding figurative components.
Finally, Ren et al.'s classification could not be used to describe chart components beyond annotation.
While these classifications informed our analysis, we struggled to use them as a means to codify the designers' intentions or goals. 
We therefore combined and extended the aforementioned classifications, resulting in a new classification having categories that explicitly acknowledge each visualization element's role.
\begin{table}[t]
    \begin{center}
    \caption{Our classification of 67 visualization thumbnails. Filled values reflect modifications from a visualization found in the article. 
    An interactive version of this table can be accessed at \url{http://hci.unist.ac.kr/SurveyVisThumbnails/}}
    \includegraphics[width=0.85\columnwidth]{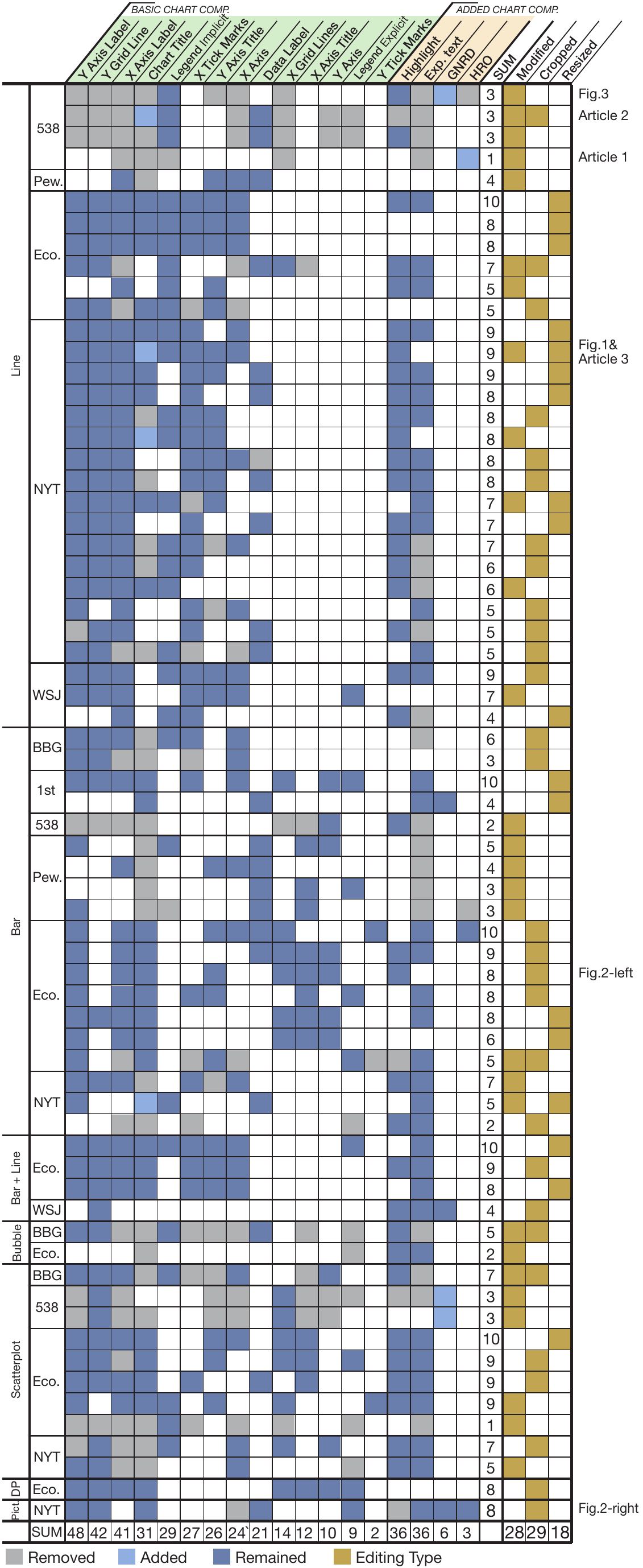}
	\label{tbl_in_the_wild_survey}
	\end{center}
\end{table}

\begin{figure}[t]   
    \centering
    \includegraphics[width=0.9\columnwidth]{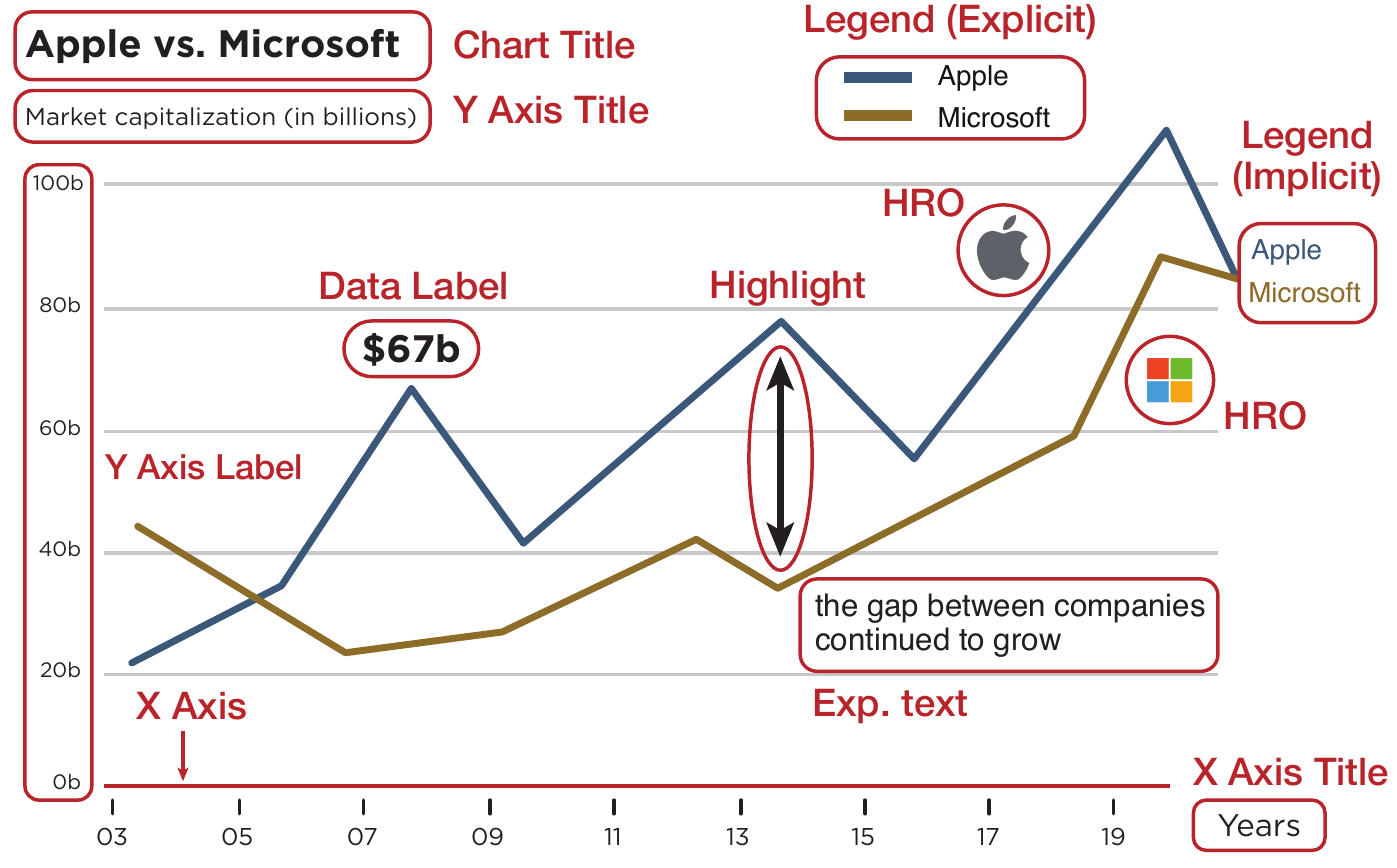}
	\caption{\textbf{Line chart annotated (in red) according to our classification of chart elements.}
  	 The chart features examples of \textsl{additional} components including HROs and highlights, as well as basic components such as axes, data labels, and legends.}
	\label{fig_add_comp}
\end{figure}

\begin{figure}[t]
    \centering
    \includegraphics[width=0.9\columnwidth]{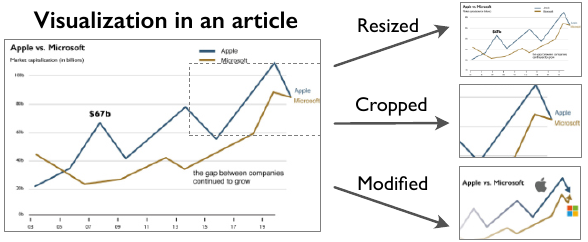}
	\caption{\textbf{Generation strategies.}
	Many news outlets generate thumbnails by resizing, cropping, and modifying original visualizations in articles.}
	\label{fig_editing_type}
\end{figure}

As a result, we identified 14 \textit{basic} and 4 \textit{[R2-18]additional} component types. 
The basic component types refer to chart components that directly facilitate chart reading and understanding (e.g., two axes in the line chart).
\autoref{fig_add_comp} shows examples of the basic component types.
They include: x- and y-axes with labels, tick marks, data labels, chart titles, and legends. 
By data labels, we mean any text that directly reflects a data value (e.g., `\$67b' in \autoref{fig_add_comp}).
We further distinguish two types of legends: \textit{explicit legends}: those drawn in a dedicated area; and \textit{implicit legends}: those drawn directly within the visualizations (e.g., Apple \& Microsoft in \autoref{fig_add_comp}). 
Additional components include explanatory text (or exp.\ text), highlights (e.g., the vertical arrow in \autoref{fig_add_comp}, the red bar in this thumbnail~\cite{Americandream}, Human Recognizable Objects (HROs)~\cite{Borkin16}), and Graphics Not Relevant to Data (GNRD) to capture all forms of graphical embellishment.
HROs are pictorial components used in legends (e.g., the Apple and Microsoft logos in \autoref{fig_add_comp} and a small human object in this thumbnail~\cite{Democrats}
or to encode data points~\cite{haroz2015isotype}.
GNRDs are images or illustrations that reflect the article's context but are not directly related the data, such as the blue image (bottom-right) in this thumbnail~\cite{Democrats}. 
\autoref{fig_editing_type} describes the current thumbnail creation strategies of news organizations (e.g., resized, cropped, and modified thumbnails).

\autoref{tbl_in_the_wild_survey}\footnote{Online version: \url{http://bitly.kr/AGRYQA}} presents the result of coding 67 visualization thumbnails\footnote{Thumbnail collection: \url{http://bitly.kr/bibRs4}} using our \textit{basic} and \textit{additional}  component classification, along with our \textit{modified} / \textit{resized} / \textit{cropped} distinction; we also indicate chart types and sources along the table's vertical axis. 
`DP' and `Pict.' in the first column (i.e., chart type) corresponding to dot plots and pictograms chart, respectively.

Our codification suggests several trends. 
Thumbnails for line charts (30 out of 67) tend to omit the X-axis title, the Y-axis, and legends.
However, they tend to include additional components such as highlights and explanation text.
Bar charts and scatterplots tend to include diverse combinations of components that are omitted or added in thumbnails. 
Example components used for the combinations include X-axis grid lines, titles, and Y-axis.
We are also able to contrast the strategies of different news organizations.
For instance, \textit{FiveThirtyEight} tends to remove nearly all components from line charts in thumbnails; they also tend to add GNRDs and HROs (e.g., \cite{Muller}).
Meanwhile, traditional print media organizations such as \textit{The New York Times} and \textit{The Economist} tend to crop or resize charts when producing thumbnails. 
Many media organizations seem to prefer modification (28) and crop (29) strategies in their designs. 
In particular, we see that FiveThirtyEight and the Pew Research Center are more inclined to modify existing visualizations, while \textit{the New York Times} seems to prefer cropping existing visualizations embedded in articles. 
Lastly, we see considerable variability in \autoref{tbl_in_the_wild_survey} which is an indication of a need for greater understanding in terms of how visualization thumbnail design choices affect readers' interpretation as well as readers' likelihood to read the article.


\begin{table}[t]
    \centering
    \caption{\textbf{Thumbnail classification.}
    24 visualization thumbnails that are not axes-based (e.g., maps) or are infrequently used.}
    \includegraphics[width=1\columnwidth]{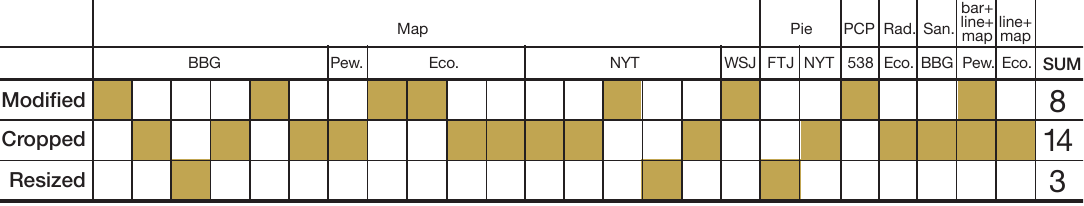}
    \vspace{-0.5cm}
	\label{tb1_non_trad_charts}
\end{table}

Table~\ref{tb1_non_trad_charts} shows the codification results of the editing strategies (i.e., modified, cropped, or resized) of 24 visualization thumbnails, including maps, pie charts, parallel coordinates, radial column charts (Rad.), Sankey diagrams (San.), and composite charts. 
We coded the editing strategies separately for the additional visualizations because those strategies can be applied to any visualizations.
As seen in Table~\ref{tb1_non_trad_charts}, cropping is more popular than the other two strategies for the additional 24 visualizations; however, cropping is not especially popular in the 67 visualizations, which include X-axis and Y-axis, as shown in Table~\ref{tbl_in_the_wild_survey}.
Most maps are presented as cropped, losing chart titles or legends.
In analyzing the thumbnails, such a removal seemed effective; it allows more space without causing additional difficulty in recognizing color-coded legends and small text. 
The discussed trends in this section are examples and by no means comprehensive.

\subsection{
Interviews with Visualization Designers
}
\label{sec_interview}

To understand the current practices in visualization thumbnail design further, we engaged in informal interviews with six visualization designers from our extended professional networks who are employed by news media organizations.
These included two journalist-engineers (with 6 and 16 years of experience, respectively), an interactive graphics developer (with 15 years of experience), a senior news article editor (with 12 years of experience), a data scientist (with 6 years of experience), and a computational journalist (with 3 years of experience).
Given the demanding nature of their work and the time zone differences between us, these interviews were asynchronous and occurred via email or within the \texttt{\#journalism} channel of the Data Visualization Society's Slack workspace.\footnote{\url{https://www.datavisualizationsociety.org}}

We asked these practitioners about two topics: (1) their intentions with respect to designing thumbnails for articles that prominently feature visualization; and (2) the challenges of incorporating visualization into article thumbnails. 
These practitioners reported a broad set of goals for thumbnails. 
First, their thumbnails must build and reinforce the organization's brand identity.
This requirement often constrains the choice of colors, font types, and HROs used in thumbnails.
Second, their thumbnails must be aesthetically appealing in order to draw readers' attention, particularly when the thumbnail appears in a visually rich social media feed.
Third, their thumbnails must also reflect any unique artwork or visual content commissioned for the article (if applicable), which might include illustrations, collages, or animations.

In regards to how they design thumbnails for data stories, these practitioners indicated that there are ``\textit{no hard and fast rules of thumb}'' that can be applied to all cases.
Aside from using a limited color palette to reinforce brand recognition, two of these practitioners admitted to avoiding the use of visualization in thumbnails.
Instead, they opted to incorporate photographic imagery whenever possible, and that photos of people appear to drive more traffic to articles.
Maps also appear to successfully drive readers to articles.  
One practitioner reported that visualizations are often regarded by readers as ``\textit{cold, intimidating, or inaccessible},'' and that while a visualization in isolation does not communicate much in itself, they can sometimes be used to accentuate a photographic or illustrated thumbnail. 
For instance, a composite thumbnail can incorporate a photo of a person with a semi-transparent visualization overlay.  
This practitioner also avoided incorporating visualization in thumbnails because they did not want to ``\textit{give away too much content}'' before the reader arrives at the article. 
Lastly, we also learned that creating a thumbnail for an article is often not the responsibility of the article author or visualization designer. 
Instead, thumbnail design is designated to designers and social media content producers who are typically not involved in writing the article.

\subsection{Visualization Thumbnail Definition and Goals}

Based on the survey results and our interviews, we define a \textit{visualization thumbnail} as a thumbnail that includes one or more visualizations.
A \textit{thumbnail}, in turn, is a small image typically between $96\times96$ and $256\times256$ pixels in dimension designed as a preview of a larger one.
Visualization thumbnails are commonly designed to be informative to allow readers to understand the gist of an article without reading, as well as enticing readers to click the article for further reading.
This does not imply that a thumbnail \emph{must} be inviting and informative to be considered a visualization thumbnail; only that if both of these design goals are satisfied, the thumbnail can be considered a \textit{well-designed} visualization thumbnail.

Two perspectives of this definition exist, and each perspective leads to different visualization thumbnail design goals.
The goal of the professional visualization thumbnail designer is to draw readers' attention to the thumbnails and to increase the click rate of the article per exposure.
This means that their design goals for a visualization thumbnail goals are not much different from those of normal thumbnail images.
Therefore, conventional design goals, such as determining an image's attractiveness~\cite{Song16} or visual aesthetics~\cite{Moere2011, Harrison15, Cawthon2007}, also apply to visualization thumbnail evaluation.

However, the goal of the consumer is to use the visualization thumbnail to select articles that best match his or her subjective criteria (e.g., preferences, interests, or intentions).
For these users, informativeness in thumbnails can be the most important requirement that allows the readers to quickly and accurately judge whether the thumbnail’s associated article meets their selection criteria.
This perspective leads to further design goals, such as informativeness, relevance, interpretability, and straightforwardness~\cite{Woodruff01, Aula10, Lam05, Li08, Teevan09}.

In an ideal scenario, the goals of producers and consumers for visualization thumbnails are the same.
However, in our interviews, practitioners with extensive experience asserted that there are cases in which they do not design visualization thumbnails to simply meet readers' needs because they also need to consider user click counts.

In this paper, we seek to understand how to design visualization thumbnails to merge both of these perspectives.
More specifically, we are looking for a middle ground where visualization thumbnails are \textbf{both} visually appealing---thus inviting more clicks---as well as representative of the underlying article---thus ensuring that readers do not feel misled.
We believe our work to be the first investigation of this topic.
Due to the limitations in the existing methods and scopes of the study, the proposed definition should be considered as a working definition and could evolve or be replaced with the results of follow-up experiments and studies. 

Though our interviews with practitioners were informal and by no means exhaustive, we were encouraged to learn about the lack of consensus in regards to guidelines for visualization thumbnail design.
We therefore remained curious about what makes visualization thumbnails effective, or how their components contribute to whether readers find them to be inviting and informative.
Conversely, we also questioned which thumbnail components could be used to mislead or misinform readers.
For example, a photo of United States special prosecutor Robert Mueller looking down may deliver a feeling of disappointment to readers.
Considered alongside our survey results, these interviews reinforced the need for further empirical study.

\begin{figure*}[t]
    \centering
    \includegraphics[width=\textwidth]{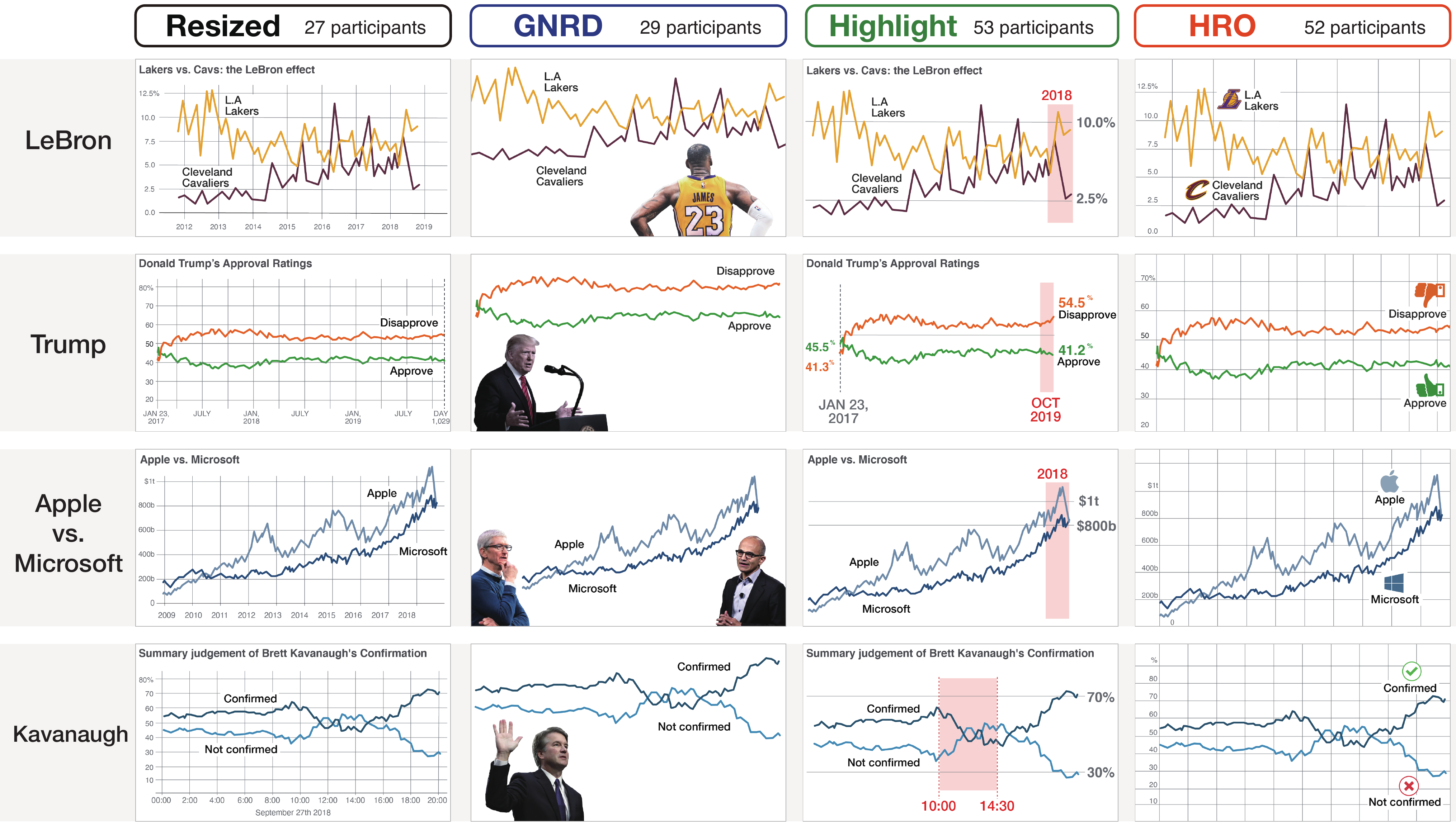}
	\caption{
 \textbf{Four types of visualization thumbnails.}
	16 thumbnails of four types with four articles from different topics that are used in our experiment. 
	Columns ($\resized{}$, $\gnrd{}$, $\highlight{}$, $\hro{}$) refer to thumbnail types, and rows (LeBron, Trump, Apple vs. Microsoft, Kavanaugh) refer to article types.
	}
	\label{fig_overview}
\end{figure*}
  
\section{User Study}
\label{sec_userStudy}

Our survey and interviews revealed a lack of studies on reader preferences for visualization thumbnail design.
To investigate these, we formulated two research questions regarding (RQ1) \textit{what design types readers would most likely click to read and why?}, and (RQ2) \textit{what do readers think of the components in visualization thumbnails?}
To answer these questions, we conducted an online user study.

\subsection{Visualization Thumbnail Design}
\label{sec_design_space}

There are three considerations for designing the thumbnails for this study, including chart and editing types and component combinations. 
In terms of chart type, we use the line chart due to its popularity (\autoref{tbl_in_the_wild_survey}) and interpretability~\cite{Lee17}.
Since we limit our discussion to the line chart thumbnails, there remains a research question on extending the design space for charts with different visualization components, such as bar, bubble, or scatterplot, and maps.
For the editing type, we exclude cropped thumbnails because they could cause unexpected issues, such as deception~\cite{Pandey15, Ritchie19, Correll20}.
We initially considered a design with many components and their combinations for RQ2, but soon found that there were too many permutations (e.g., 262,144 combinations with 18 chart components), making a user study infeasible without any guidelines for component selection. 

We therefore design our visualization thumbnails based on the chart component survey result~\cite{Kim19}, including the basic chart components. 
We also include GNRDs (Graphics Not Related to Data)~\cite{Bateman10, Borgo12, Borkin13, Borkin16, Inbar07, Hill17}, HROs (Human-Recognizable objects)~\cite{haroz2015isotype, Borkin16}, titles, and visual highlights with data labels~\cite{Hullman11, Kosara13, Lee15}, all of which affect the appeal and interpretability of a thumbnail.  

We select four articles for the study based on four considerations; thumbnails should 1) be easy for a layperson to read and understand; 2) include at least one visualization that conveys the article's main point; 
3) have a visualization for two data series that can be represented as icons or logos to help readers distinguish two or more pieces of data; 
and 4) contain the main character that can be displayed as GNRDs.
After a careful examination of candidate articles, we find four articles that satisfy our requirements. 
The topics of the four articles pertain to sports star's influence on fan bases~\cite{NewsArticle1} (LeBron), a recent U.S.\ president's approval ratings~\cite{NewsArticle2} (Trump), a comparison of Apple and Microsoft stock prices~\cite{NewsArticle3} (Apple vs.\ Microsoft), and a summary judgment of a politician's confirmation~\cite{NewsArticle4} (Kavanaugh).
Among the four articles, the articles about Trump and Apple vs.\ Microsoft contain only one visualization in each article. 
The articles about LeBron and Kavanaugh include more than two visualizations. 
When designing our visualization thumbnails for the experiment, we referred to articles' thumbnails or the visualization most discussed in the article.

Surveying visualization thumbnails with two experts in the field of design, we find that thumbnails have many design variables in the thumbnails that designers manipulate to create strong impressions of brand identity and invite more readers to click. 
These design variables include colorful backgrounds~\cite{Impeachment}, unique layouts (e.g., hiding chart areas by a figure:~\cite{Muller}), and images of faces with stimulating facial expressions and gestures (e.g., a photograph of Donald Trump:~\cite{Interest}).
Due to its large decision space and possible bias, we decide to use fixed values for the variables.
For example, we use the same color sets with original visualization, the same layouts for the same thumbnail type, and figures with neutral facial expressions. 

Based on the considerations and articles, we then present the designed visual thumbnails.  
We design 16 thumbnails of four types, as shown in \autoref{fig_overview}.
\autoref{tb3_design_space} summarizes how we control the chart component combinations.
To decide the size of the visualization thumbnails, we survey the size of visualization thumbnails from news organizations (supplementary material \url{https://osf.io/khgw2/}). 
Among different thumbnail sizes, we choose Pudding's size (width: 343px, height: 193px), because its width is close to an average (i.e., not extremely small or large), and it can sufficiently show all components of the designed thumbnails in popular mobile devices (e.g., iPhone X, display size: 375$\times$812px).

\textbf{$\resized{}$ thumbnails are the resized versions of the visualizations from the original articles.}
\textit{The New York Times} and \textit{The Economist} frequently use this design approach. 
The charts in this type tend to contain many components, including basic chart components, such as axis labels, implicit legends, and a chart title, as many other thumbnails do (e.g.,~\cite{NewsArticle3}).
There could be minor differences in axis representations (e.g., tick marks) and the use of grid lines in the original visualizations in the articles. 
For consistency with $\resized{}$ thumbnails, we use the same style of tick marks and gridlines across the thumbnails. 

\begin{table}[t]
    \centering
    \caption{\textbf{Chart components.} Identified for each thumbnail.}
    \includegraphics[width=0.9\columnwidth]{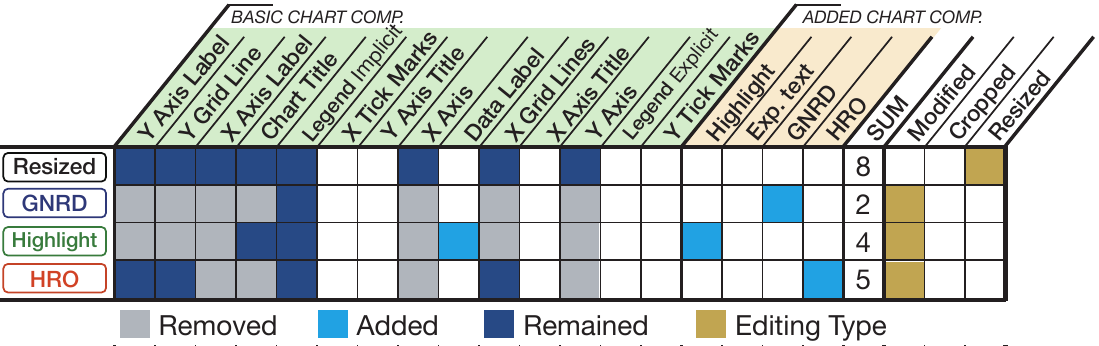}
	\label{tb3_design_space}
	\vspace{-0.7cm}
\end{table}

\textbf{$\gnrd{}$ thumbnails consist of a photograph (GNRD) and a simplified chart without any axis labels.}
GNRDs are included in the thumbnails to invite readers, stimulating their curiosity (e.g., the photography in this thumbnail of a woman walking: ~\cite{Paygap}).
They often imply information about an article’s content by showing a person who is related to the content.
In general, the chart shows an overall trend without axes, and words as implicit legends.
This type of thumbnail is frequently shown in \textit{FiveThirtyEight}'s (e.g., ~\cite{Muller}) and \textit{The Wall Street Journal}'s graphics.

\textbf{The main feature of $\highlight{}$ thumbnails is the use of a highlight block (e.g., red blocks in Fig.~\ref{fig_overview} $\highlight{}$) to emphasize part of a visualization (e.g., ~\cite{Change}).}
Data labels are added to emphasize a specific part of the data that conveys the main point of the article (e.g., `46.0\%' and `39.8\%' of this thumbnail~\cite{NewsArticle2}). 
\textit{The Wall Street Journal Graphics},\textit{FiveThirtyEight}, and \textit{Bloomberg News} frequently use this thumbnail-design approach. 
In general, the visualization in this type contains simpler axes than those in $\resized{}$ thumbnails, excluding basic components such as axis labels and gridlines. 
Reference lines can also have data labels, which provide the information of the complete y-axis (e.g., the dotted vertical line and numbers in this thumbnail~\cite{NewsArticle2}). 

\textbf{$\hro{}$ thumbnails have recognizable objects, such as logos or icons, on a plain line chart with grid lines as the background.}
Their main feature is that both logos and icons  (e.g., icons of handcuffs~\cite{Handcuff}) directly refer to the data in the visualization. (e.g., logos~\cite{NewsArticle1}). 
\textit{FiveThirtyEight} and the \textit{Pew Research Center} frequently produce this type of thumbnail. 

\subsection{Workshop with Practitioners}
\label{sec_workshop_with_practitioners}
To ensure the design quality of the designed thumbnails and to collect feedback and expectations on the thumbnail experiment, we held a workshop with 9 practitioners at a data journalism conference.
The practitioners were a data scientist (five years of experience), three graphic designers (six, four, and three years of experience, respectively), two data journalists (five and four years of experience, respectively), a computational journalist (three years of experience), and two interactive graphics developers (13 and three years of experience, respectively). 
They all worked for data journalism organizations.

As the workshop began, we described the four types of visualization thumbnails and showed the 16 thumbnails (Figure~\ref{fig_overview}) produced with associated news articles. 
We asked them to evaluate the quality of the thumbnails and their personal preferences for the thumbnail types. 
We also requested that they report any possible issues that make a thumbnail ineffective for use. 
Lastly, we asked them if they, based on their domain experience, had comments, including possible hypotheses or insights on the experiment. 
 
Overall, we received positive feedback on the thumbnails; all practitioners agreed that there were no critical issues that made the thumbnails look far different from those they had created.
For their preferences for thumbnails, seven practitioners expressed their preference for $\gnrd{}$, as they are intuitive and eye-catching: ``\textit{I can easily catch that what I would read ‘some statistics of Donald Trump’ and ‘the sales of stock prices of two companies’}."
One practitioner expressed her preference for $\gnrd{}$ thumbnails, pointing out both their interpretability and appeal:
``\textit{News articles are similar to books on a large shelf...
The eye-catching design of a book is most important.
If the book cover is informative, it will do better.}"
Meanwhile, five practitioners stated they were interested in our results because they assumed that readers could view $\resized{}$ thumbnails as unfriendly, hard to read, and cluttered, but they did not have evidence of this assumption.

To sum up, we found that the thumbnails we produced are similar to what the practitioners create. 
We also confirmed that the thumbnails could be used in the user study after revising them based on the collected design suggestions (e.g., increasing font sizes, rounding line edges).
The practitioners were also curious about the readers’ evaluation of the four types of thumbnails. 
Next, we describe how we performed the user study with the designed thumbnails. 
\begin{figure}[t]
    \centering
    \includegraphics[width=\columnwidth]{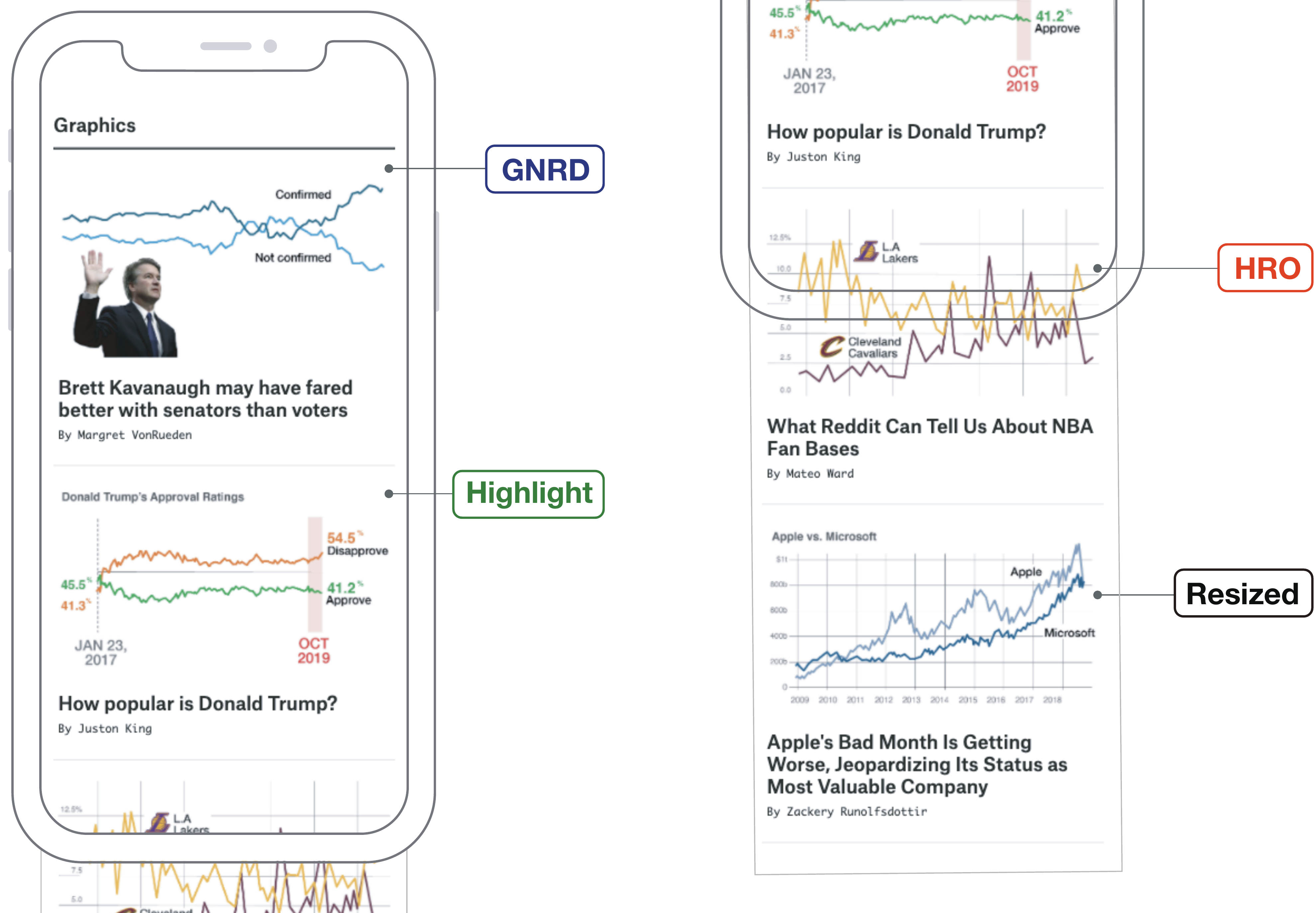}
    \caption{\textbf{Sample interface.} Mobile page used for the experiment.}
	\label{appendix_exp_mobile}
\end{figure}

\subsection{Study Procedure}
\label{sec_studyProcedure}

For the experiment, we created a web page where readers could browse news articles with thumbnails. 
For the page layout, we referred to the graphics sections of \textit{Bloomberg} (\url{https://www.bloomberg.com/graphics}) and \textit{The Wall Street Journal} (\url{https://graphics.wsj.com}).
For example, the layout presented a news article thumbnail first and then placed a title and bylines below the thumbnail. 
The web page listed four thumbnails at a time (\autoref{appendix_exp_mobile}) and allowed scrolling.

We conducted the user study with participants recruited from Prolific. 
We allowed the participants to participate in the experiment only once. 
We also prevented duplicate participation by matching their platform ID (Prolific) to a session key provided by the system to access our experiment page.
As participants entered the experiment website 
published in \textit{Prolific}, they were asked to read the introduction and electronically agree with the consent form.
They were also provided access to the IRB approval document (UNISTIRB-19-33-A).
Then, they were asked to fill out a demographic survey (e.g., gender, age, education, and frequency of using a mobile device to read news articles).
After the survey, participants were asked to read instructions that included the rules and restrictions for the experiment, such as not to create new tabs or refresh the page.

After reading the instructions, the participants entered the web page and clicked on an article that they wanted to read. 
They were asked to write a reason for their choice (a minimum of 100 characters).
Participants were encouraged to provide selection reasons for their choices and to avoid writing reasons purely based on their personal preferences on a particular topic/person/location (e.g., ``I like to read sports news more than other topics").
Finally, participants were asked a consistency check question (e.g., ``please click the article that you chose'') and then end the session.
The latter was conducted to detect any crowdworkers who were randomly clicking; since it was administered directly after the original question, this was not a memory check, but a \textit{consistency} check.
During the experiment, we recorded the participants’ device types and sizes to ensure that only mobile device users could participate.
Participants did not report any complaints about the difficulty or ambiguity of the task instructions.

We recruited 237 participants and provided the same compensation (£1.25, hourly wage rate: £9.86), regardless of the result of the consistency check. 
However, we excluded the data of 40 participants who did not show consistency in their thumbnail selection.
We also excluded the data of 36 participants who did not follow instructions for thumbnail selection and chose articles based on their personal preferences ($\resized{}$: 8, $\gnrd{}$: 12, $\highlight{}$: 9, $\hro{}$: 7). 
Specifically, we excluded reasons that only mention personal preferences without any opinion on the thumbnails. 
The excluded reasons tend to only have expressions on the topic (11 participants), article titles (4 participants), and character/company (21 participants), such as ``I chose this thumbnail because Apple is an incredibly popular company and it is very concerning when they don’t do well financially.''
We also excluded reasons where coders were unable to reach unanimous agreement due to ambiguity.
For example, we rejected this reason: ``The data seems to run parallel in a way that the other data did not.
The consistency in the data is very compelling," because coders could not agree which part of the thumbnail the statement is describing.
When participants clearly mentioned the thumbnail image as the reason for their selection, we included the data and coded it accordingly. 
A ``clear mention'' includes elements such as charts (and chart components), companies, and main characters in the article.

In total, we analyzed the responses of 161 participants (33 years old, on average, $\sigma$=9). 
They had moderate familiarity with visualizations (1 = not familiar at all, 7 = common; average score was 4) and frequently read news articles on mobile devices: more than seven times per week: 64 participants; more than five times: 42; more than three times: 40; and less than three times: 15.
During the experiment, the participants used mobile devices with an average height and width of 775.84 px ($\sigma$=101.45) and 388.40 px (min=320px, max=450px, $\sigma$=32.18), respectively.

Before proceeding with our analysis, we first tested if users differently choose thumbnails and found that participants' selections are significantly different ($p<.05$). 
We then investigated if there is the ordering effect; if the participants choose thumbnails in a specific position (e.g., always the first shown thumbnail).
For this, we calculated the probability of each position of thumbnails being selected and divided this probability into four intervals from 0 to 1 (e.g., first: $\approx0.3$, second: $\approx0.53$, third: $\approx0.8$, and last: $\approx~1$).
Using a random function that selects X from 0 to 1, the number of thumbnails selected according to random X is determined according to the probability interval described above. 
For example, if random X is 0.47, it falls into the second interval, and we suppose that the thumbnail in the second position in thumbnail combinations is selected in our simulation.
For the simulation, we used 161 sequences that are used in the user study.
We conducted this simulation 2,000 times and obtained a result that the selections were not affected by any positions of the presented thumbnails ($p < .05$).  
We uploaded the simulation code and procedure in the supplementary materials (\url{https://osf.io/khgw2/}).
We also found very little interaction between the articles and thumbnail types (Cramer's $V=0.058$) in participant choices.

\section{Results and Analysis}
\label{sec_result}

We performed statistical tests to determine the most popular thumbnail type and the reasons stated by participants for their choice (RQ1).
Then we conducted a qualitative analysis that investigated the readers' opinions on the components of the visualization thumbnails (RQ2).

\begin{table*}[t]
    \centering
    \caption{\textbf{Visualization thumbnail comment codebook.}
    Table summarizing  the selection reasons with comment counts.
    The light blue and yellow text backgrounds indicate the interpretable and inviting categories, respectively.}
    \includegraphics[width=0.85\textwidth]{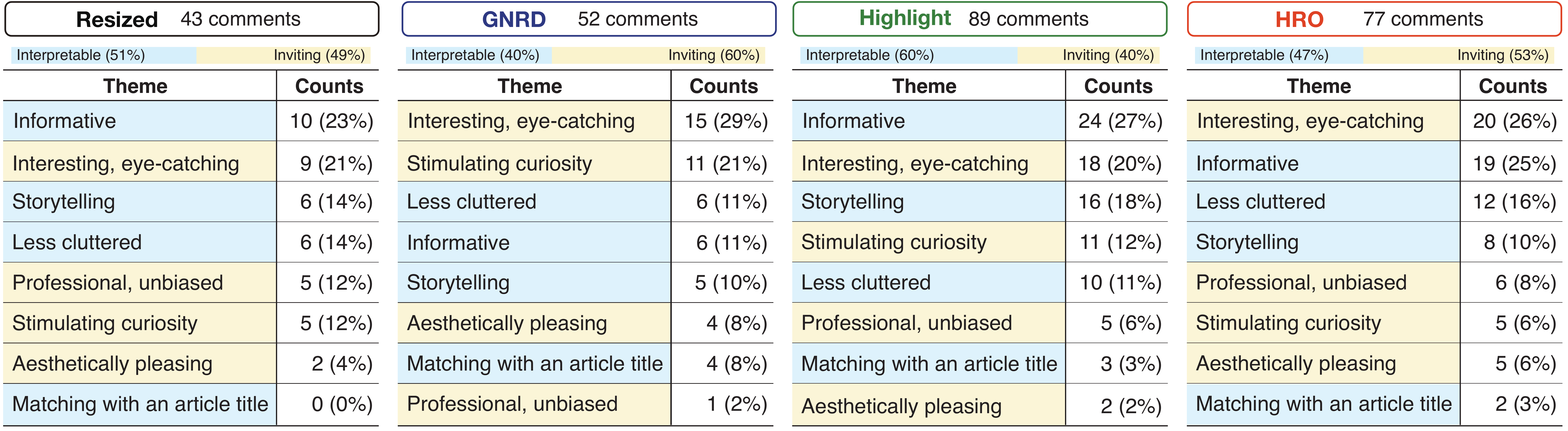}
	\label{tb5_thematic_coding}
	\vspace{-0.2cm}
\end{table*}

\subsection{Readers' Choice for Thumbnail Types (RQ1)}
\label{sec_preference}

We start by studying thumbnail type preference.

\subsubsection{Most Chosen Thumbnail Type}

First, our results showed significant preferences among the four thumbnail types ($\resized{}$: 27, $\gnrd{}$: 29, $\highlight{}$: 53, $\hro{}$: 52) according to the chi-squared test ($\chi^2$(3, N$=$161, 14.975, p$=$.0018).


To analyze the differences in users' selection among thumbnail types, we used a Thurstonian model~\cite{Yao99}, which is a scaling method for converting people's discrete selection (e.g., a vote) into physical or mental concepts (e.g., participants' tendency levels on a chosen thumbnail compared to others).
The model assumes that individual rankings are distributed around aggregate rankings. 
It also assumes that a parameter exists based on the number of votes for thumbnail $\resized{}$, $\gnrd{}$, $\highlight{}$, and $\gnrd{}$ and that the distribution of the values (i.e., preference) following Gaussian can be estimated by probability.
To quantitatively measure the preference for  thumbnail designs, we calculated the posterior distributions of their choice using Monte Carlo Markov-Chain (MCMC)~\cite{Kruschke14}, a popular method for approximating a distribution through sampling.

To visualize the distribution, the frequency distribution of the estimated thumbnail preference among the thumbnail types, we first set $\resized{}$'s frequency distribution as zero (A=0, a vertical black line in \autoref{fig_distance}) and sample 2000 times from the user selection data using the MCMC method.
\autoref{fig_distance} shows the distributions (y-axis) of the estimated thumbnail preference (x-axis) of $\gnrd{}$, $\highlight{}$, and $\hro{}$, compared to $\resized{}$.
The preference scale means that the standard deviation of the preference is 1.
So the effect size (difference of mean/standard deviation) corresponding to Cohen's d is the distance itself. 
For example, 0.1318 is the effect size.
If two distributions of $\resized{}$ and another type overlap, this means the probability of a reader choosing one of the two types is close to 50\%, respectively. 
As the overlapped part between the two distributions becomes narrow, the probability of a reader choosing $\resized{}$ becomes lower (i.e., a thumbnail types on the right side of the x-axis has a higher preference probability compared to $\resized{}$).

To find any significant difference in the preference distributions for the thumbnail types, we calculated the probability that the differences between the sampled values from any two types are positive.
When we picked two thumbnails from $\gnrd{}$ and $\highlight{}$, we found that there is a 99.5\% probability that the thumbnail from $\highlight{}$ has a higher reader preference than that from $\gnrd{}$ ($P_{(GNRD<Highlight)}$=99.5\%). 
In the end, we found that $\highlight{}$ and $\hro{}$ thumbnails have significantly higher reader preference probability distributions than those of $\resized{}$ and $\gnrd{}$ ($P_{(Resized<Highlight)}$=99.7\%, $P_{(Resized<HRO)}$=99.7\%, $P_{(GNRD<Highlight)}$=99.5\%, and $P_{(GNRD<HRO)}$=99.6\%).

\begin{figure}[t]
    \centering
    \includegraphics[width=1\columnwidth]{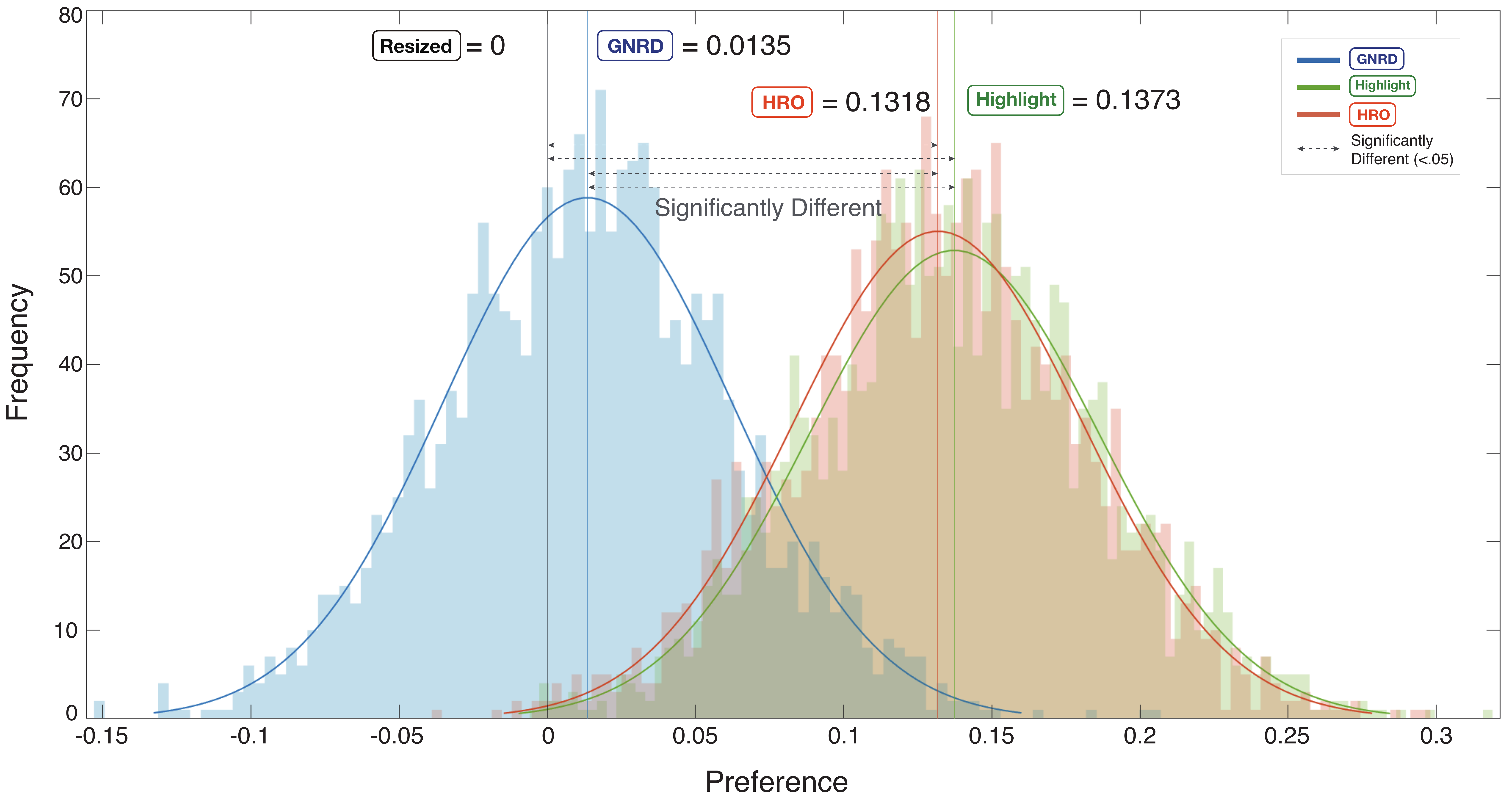}
	\caption{\textbf{Participant preferences.} 
	Posterior distribution of participants' preferences for thumbnail types shows that readers are likely to prefer $\highlight{}$ and $\hro{}$ thumbnails more than those from $\resized{}$ and $\gnrd{}$.
	}
	\label{fig_distance}
\end{figure}

\subsubsection{Thumbnail Selection Reasons}
\label{sec_qualitativeAnalysis}

To analyze the reasons for thumbnail selection, we used a bottom-up approach to thematic analysis~\cite{Braun06}.
For this analysis, the lead author of this work coded the selection reasons by developing low-level themes.
Then, two other authors reviewed the themes and revised them together to see if there is any disagreement with the proposed themes. 
Once all themes were established, the three authors independently coded the data again with the themes and later aggregated their results.
The final agreement level of the coders is 95.8\% agreement (Fleiss' Kappa = 0.26).

After building eight themes, we classified the themes into two large categories, ``inviting" and ``interpretable", which are the two main goals of visualization thumbnails (Section~\ref{sec_interview}).
\autoref{tb4_themes} summarizes the eight themes in two categories, with exemplary comments. 
For inviting categories, we included themes related to drawing readers' attention.
For example, themes related to readers' impressions of a thumbnail, such as \textit{eye-catching, professional, or aesthetically pleasing}, were classified into the inviting category. 
For the interpretable category, we included themes related to delivering articles' content to readers.
For example, \textit{less cluttered} belongs to the interpretable category because it describes the thumbnails' readability and visibility levels, which are close to the method of information delivery.
The classification resulted in 43 comments for $\resized{}$, 52 comments for $\gnrd{}$, 89 comments for $\highlight{}$, and 77 comments for $\hro{}$. 
Each participant reported 1.6 reasons for selection, on average. 

Table~\ref{tb5_thematic_coding} shows the thematic coding results, where we found that participants differently selected thumbnails with different reasons, according to a chi-squared test--
[$\resized{}$ : $\chi^2$ (7, N$=$161, 14.116, p$=$.049], [$\gnrd{}$ : $\chi^2$ (7, N$=$161, 21.231, p$=$.003], [$\highlight{}$ : $\chi^2$ (7, N$=$161, 38.191, p$=$.0001], [$\hro{}$ : $\chi^2$ (7, N$=$161, 33.026, p$=$.0001]
Specifically, readers chose $\resized{}$ and $\highlight{}$ thumbnails because they are informative, interesting, and telling stories. 
On the other hand, they picked the highlights in $\highlight{}$ thumbnails as they provide ``stimulating curiosity." 
Since the main difference between $\resized{}$ and $\highlight{}$ thumbnails are in axes, highlights, and data labels, we conjecture that the components have mainly influenced readers' final selection.
Participants chose $\gnrd{}$ and $\hro{}$ thumbnails due to their invitingness.

\begin{table}[t]
    \centering
    \caption{\textbf{Example comments.}
    Comments have been grouped into eight themes in the inviting and interpretable categories. Note that the study participants are likely paying more attention to thumbnails than an ordinary person would.}
    \includegraphics[width=\columnwidth]{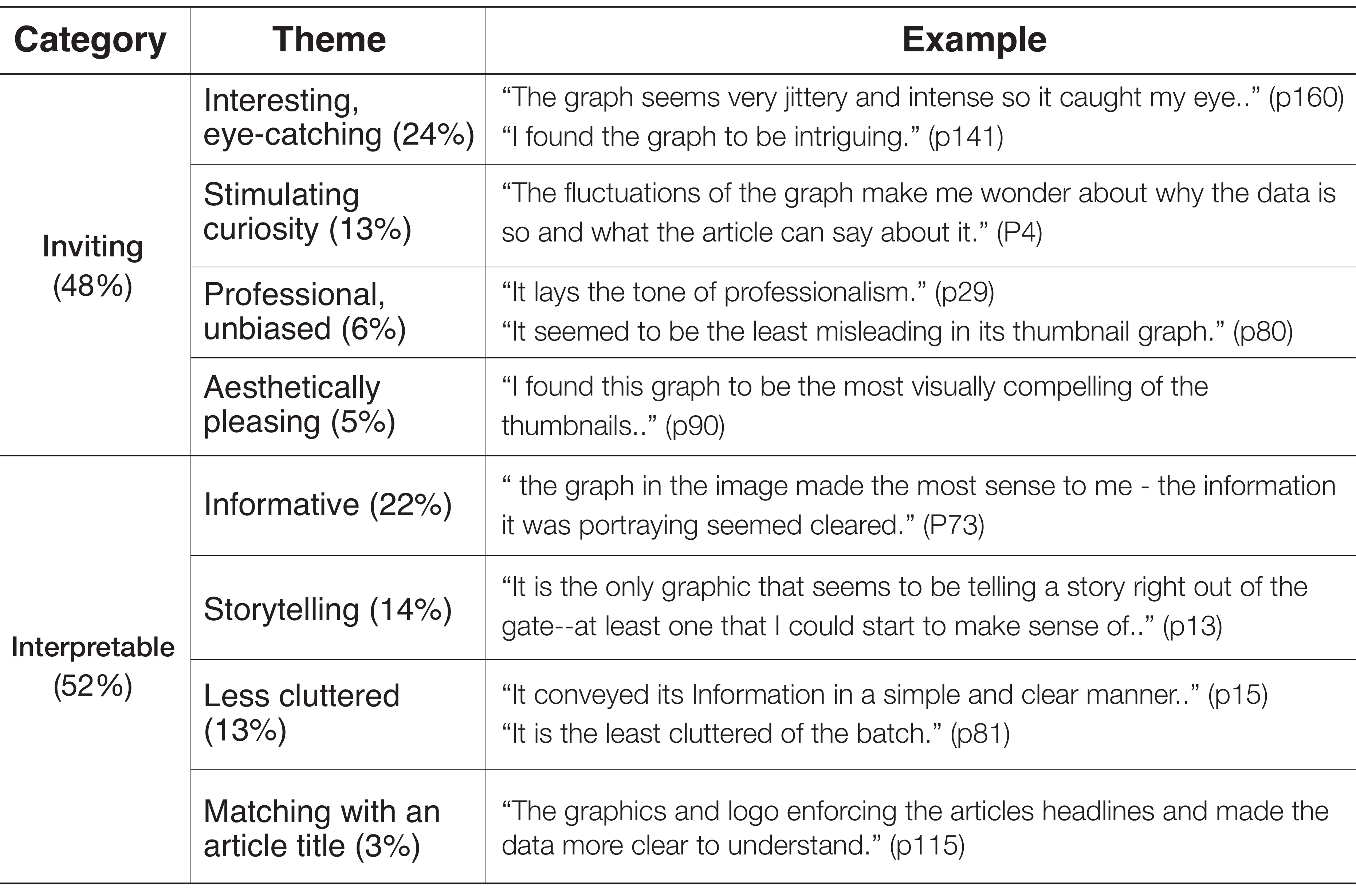}
    \label{tb4_themes}
    \vspace{-0.2cm}
\end{table}

Reviewing all analysis results, we found the reasons that readers prefer $\highlight{}$ and $\hro{}$ thumbnails over $\resized{}$ and $\gnrd{}$ thumbnails as follows.
The low preference for $\gnrd{}$ thumbnails was somewhat unexpected, as practitioners had highly rated them due to their effective communication of the articles' main ideas without causing much visual clutter (Section~\ref{sec_workshop_with_practitioners}). 
However, this result implies that the ability can be a double-edged sword, as it leads to low informativeness of the thumbnails, compared to other thumbnails, as discussed in the following section.

\subsection{Reader Opinions on Chart Components (RQ2)}
\label{sec_componentAnalysis}

We focus here on the axis information, data labels with highlights, and image components (GNRDs, HROs), as these are the primary reasons for thumbnail selection and themes.

\textbf{Abundant axes information may gain readers' trust but are rarely used for understanding data stories:}
One characteristic of $\resized{}$ thumbnails is that they are the only type of thumbnails that provide complete axes information (e.g., axis labels and values).
We first found that participants welcome the detailed axis information in $\resized{}$ thumbnails for an accurate understanding of data stories.  
Nine of 27 participants chose the axis information as to their selection reason. 
P25 stated---``\textit{The clearly visible labels on the x- and y-axis, [...] effective scale helped me quickly understand it}." 
We also noticed that thumbnails with full axes information might help readers perceive trustworthiness and professionalism from the articles (two of 27), as P19 states---``\textit{I can say that the thumbnail is my pick because it lays the tone of professionalism.}"
P25 also expressed a similar opinion---``\textit{[...] most convincing and less likely to be skewed due to the clearly visible labels on the x and y axis.}"

However, participants did not use the axis information extensively for interpretation. 
Instead, 10 of 27 participants focused on overall line chart trends (e.g., P4:  ``\textit{fluctuations of the graph make me wonder [...]}", P5: ``\textit{Apple is dropping [...]}"), which can be done quickly and easily.
These interpretations were not much different from the descriptions with $\gnrd{}$ thumbnails, which were of interest, as they had an abstract line with no axes (e.g., ``\textit{The graph seems to show more people have grown [...]}'').
These participants' thumbnail reading behavior can be viewed as a natural effort for quickly catching the main idea of thumbnails without a waste of time and in-depth analysis.
This illustrates why $\resized{}$ thumbnails do not attract readers' preference---they present too much information for the reading behavior people tend to employ for visualization thumbnails.
  
\textbf{Data labels with highlights can provide a visual guide for readers:}
Highlights~\cite{Hullman11, Kosara13, Lee15} and reference lines~\cite{Simkin87, Kong12, Hullman13} have been effective methods for attracting readers and enhancing readers' understandability of data stories with visualization. 
We confirmed the effectiveness of these techniques in thumbnails.  
For example, reference lines (e.g., the black horizontal lines of $\highlight{}$ thumbnails in \autoref{fig_overview}) tend to make participants read data stories in chronological order, working as anchors for the data labels~\cite{Simkin87, Kong12, Hullman13}.
Nine of 53 participants described how they understand the thumbnails with $\highlight{}$ thumbnails, directly referring to the data labels---``\textit{I chose this article (Article 2) [...] it was nearly a 50/50 split on January 23, 2017, and on October 19, 2019 [...]}," (P96).
Such an increased understanding on the visualization could help participants better infer the article's context, as P76 describes---``\textit{[the thumbnail] sets up the expectation the approval rating has dropped over the last two years, so I basically know the premise of the article.}"

Combining multiple components can be a good strategy to achieve both appeal and interpretability.
For example, the highlights combined with data labels can work as ``\textbf{\textit{a good visual guide}}" (P109), where highlights point where readers need to focus and data labels stress what readers should understand. 
The red blocks for presenting time ranges and data labels not only drew readers' attention to specific data points but also made readers curious about the article's context (e.g., P71: ``\textit{The article shows a \textbf{41.2\% approval rate} (in the red block), which is something I'd like to know more about, i.e., the reasons}").

\textbf{GNRD and HRO, despite being considered eye-grabbers, are associated with different keywords; GNRD with ``inspiration" and HRO with ``informativeness": }
Imagery components, such as GNRDs and HROs, were effective in engaging readers and drawing their attention~\cite{Bateman10, Hullman11}.
For $\gnrd{}$ thumbnails, 15 of 29 participants chose GNRDs as their reason for the selection, because GNRDs were ``\textit{much more visually engaging}" (P49) and ``\textit{making the thumbnail stand out more}" (P41).
Participants who chose $\hro{}$ thumbnails also described HROs' role as attention grabbers--``\textit{The inclusion of sports logos caught my eye}" (P132), ``\textit{[...] logos were recognizable and got my attention.}" (P152).
We additionally found a possibility that visualization thumbnails with GNRDs could be more effective with a title in attracting readers, which is the main access point for further reading~\cite{Kaasten02, Aula10}, as P54 stated---``\textit{The photo caught my eye, which leads to me reading the title.}"

Though both GNRDs and HROs attract readers, participants' interpretation of GNRDs and HROs may differ in visualization thumbnails.
In the study, participants with HROs tend to intuitively understand the visualization, using HROs as ``\textit{\textbf{visual legends}}" (P136). 
For example, in the $\hro{}$ thumbnail for Trump in \autoref{fig_overview}, the thumbs up or thumbs down icons represent approval or disapproval, which are easy to understand as they are consistent with text labels (i.e., `approve' and `disapprove')---``\textit{Thumbs up and thumbs down is a \textbf{universally used graphic representation} which was easy to understand for me.}"
This supports previous work stating that HROs can convey representative meanings in the form of icons or logos but also create data redundancy by conveying meaning combined with text~\cite{Borkin16}.

For GNRDs, incorporating human photos in charts could be inspiring but may lead to open interpretations.
Three participants (3 of 29) believed that GNRDs effectively informed the article's main character--``\textit{This one stood out because it had a picture of a person mentioned in the article [...] (with the GNRD) easier to think about because you don’t need to visualize the person mentioned.}"
We also observed that the meanings of human photos might be accepted differently by readers.
For example, with the same thumbnail ($\gnrd{}$ thumbnail for the stock prices), P31 commented that ``\textit{(The people in the thumbnail) are selling Apple stock}", while P35 reported that ``\textit{[...] they were talking about and what they determine is a bad month for Apple [...]}". 
We think this can be aligned with the previous finding that thumbnails with human photographs construe interpersonal and textual meanings~\cite{Knox09Punctuating}.

We conjecture that informativeness of chart components is another important factor that affects the final selection of thumbnails. 
For example, the participants may consider $\hro{}$ thumbnails to be both attractive and informative, delivering the meaning of thumbnails in a straightforward manner with text labels.
GNRDs also attract readers' attention, but we find the possibility that the participants could interpret the meaning of GNRDs somewhat differently. 
However, given the small number of participants mentioning the difference between GNRDs and HROs, there may be more as-yet unknown effects of GNRDs. 

\section{Discussion}

We have shown that readers want visualization thumbnails that not only keep their eyes focused on the thumbnail (i.e., inviting), but also convey the article's main point (informative).
Our findings indicate that thumbnail design can be aided by understanding chart component roles. 

\subsection{Lessons Learned and Design Implications}
\label{sec_DesignImplications}

Figure~\ref{fig_diagram} summarizes the role of chart components for visualization thumbnails. 
We place GNRDs, HROs, and highlights in the inviting category, while we place axes, implicit legends, and data labels in the interpretable category. 
We also find that data labels with highlights and HROs with implicit legends can work for both inviting and interpretable in the visualization thumbnails. 
We acknowledge that the participants in our study are more likely to pay attention to thumbnails than the general population.
Here we provide design suggestions that consider the roles of the chart components and the combination of these components.

\begin{figure}[t]
    \centering
    \includegraphics[width=0.8\columnwidth]{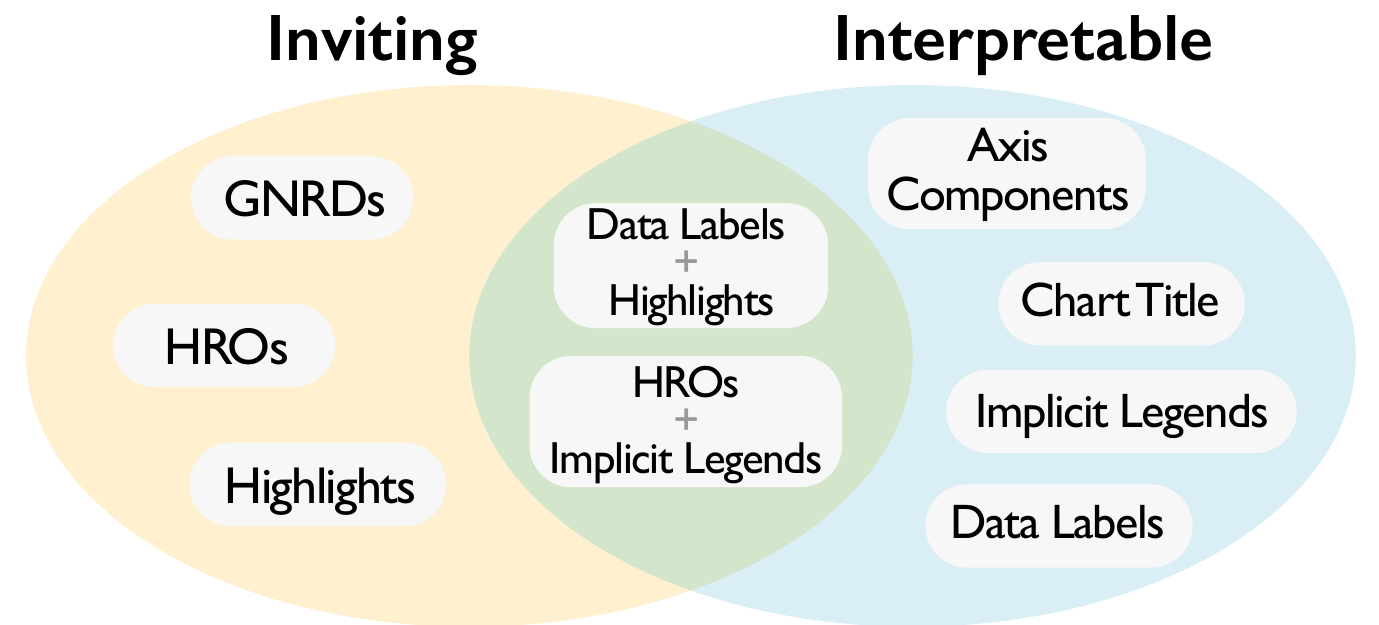}
	\caption{\textbf{Visualization thumbnail roles.}
	Interpretability and invitation.}
	\label{fig_diagram}
\end{figure}

\textbf{Lesson 1: Readers want easy-to-understand visualization thumbnails.}
As thumbnails present stories, readers expect to quickly and easily grasp the key information of the story by viewing the thumbnails.
As such, we think that summarized and simple charts in thumbnails are preferable compared to detailed charts, as shown in our results; readers click thumbnails with brief summaries ($\highlight{}$, 58) rather than full charts for the articles ($\resized{}$, 27).  
We additionally think that explicit and concise keywords can play an important role in presenting the main point of the story, while maintaining simple visualization thumbnails---``\textit{The graph in the thumbnail is easier to look at. 
The large text makes it a lot clearer what it's supposed to be about at a glance compared to the others},'' as P72 states.
Chart axes can aid readers in understanding the thumbnail, but they may not be effective if readers do not award them sufficient attention. 

\textit{\textbf{Suggestion 1:} Including descriptive text with a summarized visualization can make a thumbnail more understandable.}
We notice that chart titles and data labels help readers to grasp an article's key messages, making a thumbnail more interpretable (blue part in \autoref{fig_diagram}).
\textbf{Chart titles} should summarize the entire chart to help readers get a quick and intuitive understanding on the thumbnails.
\textbf{Data labels} of numbers can be critical design points that help readers access the detailed but concise information of the article.

\textbf{Lesson 2: Readers want attractive visualization thumbnails.}
Conventionally, eye-catching has been an important factor in thumbnail design~\cite{Topkara12, Matejka13, Song16, Knox09, Knox09Punctuating}. 
We confirm in this work that readers also expect to see a thumbnail that is attractive enough to draw their attention and invite them for further clicks. 
In particular, we find that readers prefer thumbnails with attractive chart components. 
Many readers first catch images or highlights in thumbnails and then become interested enough to explore the data and read the article---``\textit{The red block made the thumbnail look more entertaining to learn about}" (P92).

\textit{\textbf{Suggestion 2:} Including highlights or visual embellishments can make a visualization thumbnail engaging.}
Incorporating highlights or images in thumbnails is a promising strategy for drawing the reader's attention to the thumbnail (i.e., the yellow part of \autoref{fig_diagram}).
\textbf{Highlights} make a part of the chart stand out among other components and draw readers' attention to the story.
Visual embellishments such as GNRDs and HROs can be also used for eye-catching purposes, but we suggest that using them with cautions because it is possible that they can mislead readers. 
For example, portrait photography (GNRDs) can bring unexpected meaning when interpreted with given chart in a thumbnail.

\subsection{Limitations and Open Research Areas}

Although revealing readers' thoughts on inviting and interpretable visualization thumbnails throughout the user study are the first attempt at visualization thumbnail design, there are limitations in this work. 
In this section, we discuss the limitations of this work and present future research questions beyond this work. 

\textbf{Exploring the role of chart components from qualitative analysis:}
We investigated readers' opinion on chart components and their roles (RQ2) providing other aspects of visualization thumbnails that previous studies did not point out.
For example, we suggested the different roles of HROs and GNRDs in visualization thumbnails which have not been considered thus far. 
Though the obtained results are meaningful in the sense that they are the first responses extracted from the readers, the result on the effectiveness and interpretation of the chart components should not be considered as definitive yet due to the limited number of participants and extrapolations.
We think a future study could use the findings in this work in selecting chart components to verify the specific role of the selected components and their combinations.
In this work, we focus on exploring the role and effectiveness of different components in thumbnails. 
We think that a future study could investigate the role of the components with consideration of topic, person, and location perspectives.

\textbf{Confirming prior work: }
As the first work for investigating readers' preference and thoughts on visualization thumbnail designs, we provided a series of findings, and insights, and lessons from different analytical perspectives.  
However, when compared to prior work, our findings may be considered limited in that mostly confirm existing findings.
We think a future study may additionally validate our hypotheses by focusing on the differences among other types of thumbnails with different modification strategies based on our work.

\textbf{Different visualization thumbnail sizes:}
Different readers may have different thumbnail size preferences and familiarity.
As our work is the first study on visualization thumbnail design, we used a single thumbnail size based on our survey.
Thus, the effect of thumbnail size remains unanswered.
A future study may investigate the possible effects of different (smaller) visualization thumbnail sizes on different computing platforms.
This may implies there is a space for testing various thumbnail sizes including different size of components and fonts, as we described the roles of effect of axes in Section~\ref{sec_componentAnalysis}.
Our survey results show a variety of sizes being used in online news media. 

\textbf{Presenting one message vs.\ summarizing a full story:} 
Designers can choose a message presentation strategy between conveying one fact/perspective or summarizing a data story’s full narrative with a visualization thumbnail. 
Among the visualization thumbnails used in the experiments, we conjecture that the $\highlight{}$ thumbnails seem to support the story summary strategy by presenting a baseline story with the time-series data and exposing the main point with the red highlight and data label.
This can be related to picture superiority effect, in which image content is more likely to lead to an increased understanding of the material~\cite{Childers84, Shepard67, Mcbride02, Whitehouse06, Ally09, Curran11, Yangandul18}.
As we did not consider the message presentation strategy in our thumbnail design, there exist many questions on the matter. 
For example, putting multiple or collage visualizations (e.g., bar+line, bar, and line) in one visualization thumbnail can also be considered as a message presentation strategy, a topic that can be further explored in future studies. 

\textbf{Design space generalization:}
In this work, we designed and conducted a user study with visualization thumbnails that present line charts inside them. 
As such, some insights derived from this work may not be directly applicable to other types of charts (e.g., maps) with different component sizes, such as font sizes.
For example, we described the roles and effect of axes in Section~\ref{sec_componentAnalysis}, but further research is needed to investigate whether the same effect occurs with other chart types, such as bar chart, bubble chart, or scatterplots.
We used two editing types in this work---resized and modified thumbnails---but a future study could investigate other modification strategies, such as cropped thumbnails.
Additional experiments can be performed for investigating other possible impacts, such as the impacts of interactions (e.g., scrolling) and surrounding colors. 
Lastly, we used two time-series in this study because we wanted to understand how readers understand implicit labels and HROs in situations involving more than one time-series.
We were concerned that using more than two time-series could lead to overly complicated experiments given the lack of prior work or guidelines.
Nevertheless, conducting experiments with more than two time-series is an interesting extension for future work, and could shed light on additional effective thumbnail designs.

\textbf{Different highlighting techniques: }
To investigate the impact of the presence of highlights in visualization thumbnails, we chose highlights that are frequently seen in the thumbnail designs (e.g., red block, reference lines).
However, other highlighting techniques could show better performance. 
Examples include changing opacity or colors~\cite{Kong12, Ren17} and adding graphical annotations (e.g., circles) ~\cite{Kong12, Satyanarayan14}.
Future work should investigate the role of the highlight techniques on readers' perception.

\textbf{Impact of stimulating portrait photography (GNRDs):}
In this work, we used photos that have neutral impressions and gestures in an effort to prevent any possible biases from any other features related to charts (Section \ref{sec_design_space}).
With the photos, we find that portrait photography not only catches readers' attention but also tells a story when used in visualization thumbnails, extending the previous works \cite{Caple12, Barrett05, Knox09, Knox09Punctuating}.
However, this result might not apply to all other thumbnails, as there are many design choices for photo selection in real-world thumbnails(e.g., a smiling or frowning face). 
Future work could investigate the impact of stimulating photos (GNRDs) with different features. 
For example, we can study readers' attention on various face expressions of photography in visualization thumbnails, compared to other components, such as headline texts.

\textbf{Measuring a visualization thumbnail's helpfulness after reading an article:}
In this work, we asked readers to choose an article to read solely based on thumbnails. 
Thus, we did not measure the perceived usefulness of visualization thumbnails. 
Measuring the perceived helpfulness~\cite{Aula10} of visualization thumbnails is an important issue, as it determines news organizations' trust and reliability.
For example, after a reader reads an article, but the content or conclusion of the article is not what is expected from the thumbnail, the reader may skip thumbnails from the news organizations. 
But what kind of factors or components involves the usefulness of the thumbnails is not clear. 
Our study indicates that GNRDs are one candidate that can affect helpfulness of thumbnails, as they can be either informative or misleading.  
As such, a future study needs to investigate the perceived usefulness of visualization thumbnails and the factors involved with the usefulness.

\textbf{Impact of aesthetically-pleasing, artistic visualization thumbnails: } 
The visual aesthetics of visualization have been discussed from many perspectives, including subjective impressions of visualization and design criteria~\cite{Cawthon2007, Harrison15}.
In our study, the practitioners stated that they put much effort on producing visually appealing thumbnails (\autoref{sec_interview}).
We also observed many artistic visualization thumbnails from news organizations. 
For example, \textit{FiveThirtyEight}, \textit{the Wall Street Journal Graphics}, and \textit{the Pudding} tend to adjust the color and contrast of photos (GNRDs) and incorporate them into thumbnails together.
However, due to the larger design space and lack of clear criteria for designing visually pleasing visualization thumbnails, we did not discuss the topic in this work. 
For example, to design a visually pleasing visualization thumbnail, one may need to consider how to control biases from colorful backgrounds, unique layouts, and face photos with different facial expressions and gestures.
Future research could investigate the aesthetics design space of visualization thumbnails and answer what makes visualization thumbnails visually appealing to readers.

\textbf{Honesty of responses:}
In the experiment, we followed current best practice in crowdsourcing---we paid participants a sufficient and fair amount of compensation and did not give any systemic clue that could lead them to certain answers.
However, we cannot be sure of the honesty of participant responses.
On the other hand, we have not found any indication that participants lied about their answers. 
Furthermore, we believe that in our experimental design, honesty is less effort than dishonesty for participants because that means they don’t have to put effort into fabricating dishonest answers and reasons.

\textbf{Potential ordering effects:}
We used randomization rather than counterbalancing, so there is a potential of an ordering effect impacting our findings. 
To find any possible ordering effect on thumbnail or article presentation orders, we conducted a user thumbnail selection simulation 2,000 times as described in Section~\ref{sec_studyProcedure}, where participants choose thumbnails based on the selection probability in the study. 
Thumbnails can be chosen in any position during the simulations.
Our findings from the simulation show that participants were significantly unlikely  to be affected by thumbnail positions.


\section{Conclusion}

We began this project by asking what makes thumbnails for data stories inviting and interpretable.
We surveyed visualization thumbnails and had a series of interviews with practitioners about the design of thumbnails for data-driven stories.
Based on our survey, we defined the design types of visualization thumbnails and conducted a user study to determine the most appealing thumbnail design.
Our study results reveal a design space for thumbnails: a set of thumbnail design guidelines that can be leveraged to attract readers and help them understand the context of articles associated with thumbnails.
The results also indicate that chart components are the keys in visualization thumbnails to attract readers’ attention and enhance readers' understandability of the visualization thumbnails.
We also report various thumbnail design strategies by effectively combining the chart components, such as a data summary with highlights and data labels and a visual legend with text labels and HROs.
Ultimately, our study sheds light on an uncharted design space for visualization thumbnail design and toward automatically generating or recommending an ideal set of visualization components to include in a thumbnail.

\section*{Acknowledgments}
This work was supported by the Korean National Research Foundation (NRF) grant (No. 2021R1A2C1004542), by a grant of the Korea Health Technology R\&D Project through the Korea Health Industry Development Institute (KHIDI), funded by the Ministry of Health \& Welfare, Republic of Korea (grant number:HI22C0646), and by the Institute of Information \& Communications Technology Planning\&Evaluation (IITP) grants (No.2020-0-01336–Artificial Intelligence Graduate School Program, UNIST), funded by the Korea government (MSIT). 

\bibliographystyle{IEEEtran}
\bibliography{thumbvis}

\begin{IEEEbiographynophoto}{Hwiyeon Kim} received her B.S.\ and M.S.\ degrees from UNIST (Ulsan National Institute of Science and Technology) in Ulsan, South Korea in 2019 and 2021.
Her research interests include data journalism and HCI. 
\end{IEEEbiographynophoto}

\begin{IEEEbiographynophoto}{Joohee Kim}
received her B.S.\ degree in Computer Science and Engineering from UNIST (Ulsan National Institute of Science and Technology) in Ulsan, South Korea.
She is working toward her M.S.\ degree at UNIST.
Her research interests include data visualization and HCI. 
\end{IEEEbiographynophoto}

\begin{IEEEbiographynophoto}{Yunha Han} received her M.S.\ in 2021 from UNIST (Ulsan National Institute of Science and Technology) in Ulsan, South Korea. 
She is a data engineer at NCSoft.
Her research interests include data visualization and HCI. 
\end{IEEEbiographynophoto}

\begin{IEEEbiographynophoto}{Hwajung Hong} received the Ph.D.\ degree from Georgia Institute of Technology in 2015.
She is an associate professor in the Department of Industrial Design at KAIST.
Her research lies at the intersection of human-computer interaction and journalism.
Her research interests include the social implications of data and artificial intelligence.
\end{IEEEbiographynophoto}

\begin{IEEEbiographynophoto}{Oh-Sang Kwon}
received the doctoral degree in psychological sciences at Purdue University in 2009.
He is an associate professor in the department of biomedical engineering at UNIST (Ulsan National Institute of Science and Technology) in Ulsan, South Korea.
His research interests include human visual perception, perceptual/cognitive biases, perception-action interaction, and perceptual learning.
\end{IEEEbiographynophoto}

\begin{IEEEbiographynophoto}{Young-Woo Park} received the Ph.D.\ degree in industrial design in 2014 from KAIST in Daejeon, South Korea. 
He is an associate professor of design at UNIST (Ulsan National Institute of Science and Technology) in Ulsan, South Korea.
As a director of the Interactive Product Design Laboratory, his research highlights the significance of 'physical richness' during interaction with technologies.
He explores the potential of embodiment of tactile channel, shape-changing and data materialization as a method for designing aesthetic interaction. 
\end{IEEEbiographynophoto}

\begin{IEEEbiographynophoto}{Niklas Elmqvist}
received the Ph.D.\ degree in 2006 from Chalmers University of Technology in G\"{o}teborg, Sweden.
He is a full professor in the College of Information Studies at University of Maryland, College Park in College Park, Maryland.
He is also a member of the Institute for Advanced Computer Studies (UMIACS) and formerly director of the Human-Computer Interaction Laboratory (HCIL) at University of Maryland.
He is a senior member of the IEEE and the IEEE Computer Society.
\end{IEEEbiographynophoto}

\begin{IEEEbiographynophoto}{Sungahn Ko}
received the doctoral degree in electrical and computer engineering in 2014 from Purdue University in West Lafayette, IN, USA.
He is an associate professor in the school of Computer Science and Engineering at UNIST (Ulsan National Institute of Science and Technology) in Ulsan, South Korea.
His research interests include visual analytics, information visualization, and Human-Computer Interaction.
\end{IEEEbiographynophoto}

\begin{IEEEbiographynophoto}{Bum Chul Kwon}
received his M.S.\ and Ph.D.\ in Industrial Engineering from Purdue University in 2010 and 2013, and his B.S.\ in Systems Engineering from University of Virginia in 2008.
He is a research scientist at IBM Research.
His research area includes visual analytics, data visualization, human-computer interaction, healthcare, and machine learning.
Prior to joining IBM Research, he worked as postdoctoral researcher at University of Konstanz, Germany.
\end{IEEEbiographynophoto}

\vfill

\end{document}